\documentclass[12pt,letterpaper]{article}  

\usepackage[includeheadfoot,
            marginratio={1:1,2:3}, 
            width=460pt, 
            height=688pt,]{geometry}

\usepackage{physics}
\usepackage{amsmath}
\usepackage{amsfonts}
\usepackage{amssymb}
\usepackage{mathtools}
\usepackage{graphicx}
\usepackage{tikz}
\usepackage{tikz,tcolorbox}
\usepackage{booktabs}
\usepackage{bbding}
\usepackage{pifont}
\usepackage{adjustbox}
\usepackage[utf8]{inputenc}
\usepackage[splitrule]{footmisc} 
\usepackage{slashed}
\usepackage{float}

\usepackage{empheq}
\usepackage{paralist}
\usepackage{cite}
\usepackage[hidelinks]{hyperref}
\usepackage{xcolor}
\usepackage{tensor}

\definecolor{darkblue}{RGB}{0,0,200}   %
\newcommand{\nc}{\newcommand}
\nc{\lb}{\llbracket}
\nc{\rb}{\rrbracket}
\nc{\gl}{\llbracket}
\nc{\gr}{\rrbracket}
\nc{\del}{\partial}

\nc{\eq}[1]{\begin{equation}
                     \begin{split} #1 \end{split}
                     \end{equation}}
\nc{\ul}{\underline}
\nc{\ov}{\overline}

\nc{\fa}{\hat}
\nc{\fb}{\MakeUppercase}
\nc{\fc}{\tilde}

\allowdisplaybreaks[2]
\numberwithin{equation}{section}

\begin{document}
\renewcommand*{\thefootnote}{\fnsymbol{footnote}}
\hypersetup{pdfborder={0 0 0}}

\vspace*{-1.5cm}
\begin{flushright}
  {\small
  MPP-2026-25
  }
\end{flushright}
\vspace{50pt}
\begin{center}
{\LARGE A domain wall bound on anti-de Sitter vacua}
\end{center}
\vspace{0.1cm}

\begin{center}
{Niccol\`o Cribiori$^a$, Antonia Paraskevopoulou$^b$, Thomas Van Riet$^a$}
\end{center}

\vspace{0.1cm}
\begin{center} 
\emph{
$^a$ KU Leuven, Institute for Theoretical Physics,\\ Celestijnenlaan 200D, B-3001 Leuven, Belgium \\
 \vspace{0.3cm}
 $^b$ Max-Planck-Institut f\"ur Physik (Werner-Heisenberg-Institut), \\[.1cm] 
   Boltzmannstra\ss e 8,  85748 Garching, Germany\\
    \vspace{0.3cm}
    } 
\end{center}

\vspace{0.5cm}

\begin{abstract}

We consider anti-de Sitter flux vacua interpolated by flux-changing domain walls. Demanding that the tension of such a domain wall be above the ultraviolet cutoff of the effective description, we derive an upper bound on the anti-de Sitter radius, which we term domain wall bound. It translates into a lower bound on the gravitino mass, thus realizing the gravitino conjecture and the anti-de Sitter distance conjecture of the swampland program.
We test the domain wall bound on several examples with a candidate hierarchy of scales: classical flux vacua, racetrack models, LVS and KKLT-like anti-de Sitter vacua.  The classical flux vacua and LVS are found to be compatible with the bound. For racetrack and KKLT-like anti-de Sitter vacua, the bound poses a non-trivial constraint on achieving large hierarchies of scales.
\end{abstract}

\thispagestyle{empty}
\clearpage
\renewcommand{\thefootnote}{\arabic{footnote}}
\setcounter{tocdepth}{2}

\tableofcontents
\newpage

\section{Introduction}

Explaining the large hierarchies of scales observed in the universe is among the central challenges of any microscopic description of gravity. In string theory the problem is particularly interesting since the theory is known to mix ultraviolet (UV) and infrared (IR) scales in a non-trivial way, see $e.g.$~\cite{Kutasov:1990sv}. One may therefore expect that consistent effective descriptions of gravity also feature some form of UV/IR mixing at low energy. Among the simplest possibilities is the fact that the UV and IR cutoffs might not be independent of one another. In this work, we investigate a specific realization of such phenomenon.

One way to arrive at a relation of this sort is to argue in terms of entropy bounds; this is perhaps the most explored route. 
At least heuristically, entropy bounds, such as the covariant entropy bound conjecture \cite{Bousso:1999xy}, capture the idea that the entropy of a certain spacetime region cannot exceed a quarter of the area bounding the region and measured in Planck units \cite{Bekenstein:1980jp,Fischler:1998st}. 
As such, they are a manifestation of holography \cite{Susskind:1998dq}. The covariant entropy bound can be conveniently phrased in anti-de Sitter (AdS) spacetime, especially in the absence of black holes, as the  spacelike projection theorem applies\cite{Bousso:1999cb}. This permits focusing on spacelike regions rather than null hypersurfaces, the lightsheets, on which it was originally formulated \cite{Bousso:1999xy}. Under some reasonable assumptions, one can then deduce an upper bound on the UV cutoff in terms of the IR cutoff and the Planck scale, investigated recently in \cite{Castellano:2021mmx,Calderon-Infante:2023ler,Herraez:2024kux,Andriot:2025cyi,Cribiori:2025oek} and given below in \eqref{eq:CEBd}.

In the present work, we explore a second way to arrive at a relation of this sort. It exploits the existence of fundamental domain walls interpolating between two different phases of the effective theory, and has been recently put forward in \cite{Cribiori:2024jwq} in the context of two-dimensional AdS flux vacua. Here, we focus instead on AdS vacua in more than two spacetime dimensions. It is perhaps \emph{a priori} unexpected that domain walls and entropy bounds contain similar physical information, but this is what our analysis suggests.

We start by discussing the notion of IR cutoff in AdS, addressing related subtleties, and then we provide a derivation of the aforementioned bound on the UV cutoff from the equations of motion of ten-dimensional supergravity, under some assumptions. Due to the role played by fundamental domain walls in our derivation, we will refer to this bound as \emph{domain wall bound}. As byproduct, we also establish the validity of the gravitino conjecture \cite{Cribiori:2021gbf,Castellano:2021yye} and the anti-de Sitter distance conjecture \cite{Lust:2019zwm}, under the same assumptions. We discuss the limitations of our top-down derivation and we test the domain wall bound in different models in the literature of string compactifications. 
Our strategy will be as follows: we work at the level of the $d$-dimensional effective theory, which we assume to exist, and run our analysis until we reach a contradiction, if that happens.

Our rationale behind the chosen examples is really the fact that the domain wall bound bounds large hierarchies of scales. Indeed, it puts a constraint on the UV and IR cutoffs that is not obvious from a low-energy perspective, and that is complementary to the standard requirement of asking that the former be larger than the latter. Since the UV and IR cutoffs are respectively the maximal and the minimal scale in an effective theory, bounding them means bounding any hierarchy in the between. With this motivation, we look at supergravity vacua featuring scale separation. These can be purely classical, as those related to the DGKT construction \cite{DeWolfe:2005uu,Camara:2005dc} in four dimensions, and their three-dimensional counterparts \cite{Farakos:2020phe}, or they can rely on the existence of suitable quantum corrections, as those along the lines of the  KKLT \cite{Kachru:2003aw} or the Large Volume Scenario (LVS) \cite{Balasubramanian:2005zx}. Since KKLT-type scenarios are based on exponentially small gravitino mass, they can violate the domain wall bound parametrically, whereas DGKT-type and LVS-type constructions satisfy the bound. We then discuss how to interpret this result and finally we draw our conclusions and discuss future directions.

\section{A domain wall bound in AdS}

We consider a $d$-dimensional, weakly coupled effective field theory (EFT) in AdS with $N$ light degrees of freedom up to the UV cutoff $\Lambda_{\rm UV}$ and above the IR cutoff $\Lambda_{\rm IR}$. Placing the system in a sphere of radius $\Lambda_{\rm IR}^{-1}$, the covariant entropy bound constrains the extensive EFT entropy as $S_{\rm EFT} = N (\Lambda_{\rm UV}/\Lambda_{\rm IR})^{d-1} < \frac 14  (M_{\rm Pl,d}/\Lambda_{\rm IR})^{d-2}$, where $M_{\rm Pl,d}$ is the $d$-dimensional Planck mass. For our purposes, the bound is thus represented by 
\begin{equation}
\label{eq:CEBd}
\Lambda_{\rm UV}^{d-1} \leq  M_{\rm Pl,d}^{d-2} \Lambda_{\rm IR}\,,
\end{equation}
which we assume valid up to order-one coefficients.
As discussed $e.g.$~in \cite{Castellano:2021mmx,Calderon-Infante:2023ler,Herraez:2024kux,Andriot:2025cyi,Cribiori:2025oek}, this relation is a form of UV/IR mixing which is to be expected in holography and quantum gravity. 
Exploiting basic properties of fundamental domain walls, such as their tension, we will provide a top-down derivation from string theory of a bound closely related  to \eqref{eq:CEBd}.

Indeed, a central issue in \eqref{eq:CEBd} is to correctly identify the UV and IR cutoff scales. For the former, we will discuss our choices, and the motivation behind them, when looking at specific examples. For the latter, we present now certain arguments in favor of employing the AdS radius, $L_{\rm AdS}$. Ultimately, this choice will be supported by our top-down derivation in section \ref{sec:derbound}.

\subsection{AdS and the IR cutoff}

The main property of AdS of interest to us is that it acts as IR regulator, despite having infinite volume.  The intuition is that interactions change behavior when probing distances larger than AdS radius $L_{\rm AdS}$, as originally described in \cite{Hawking:1982dh,Callan:1989em}.

First, let us consider electromagnetic interactions. From the AdS metric in global coordinates,
\begin{equation}
ds^2 = -\cosh^2(r/L_{\rm AdS}) dt^2 + dr^2 + L_{\rm AdS}^2 \sinh^2(r/L_{\rm AdS})d\Omega_{d-2}^2\, ,
\end{equation}
we see that slices of constant time, $dt=0$, are hyperbolic. Hence, the area $A$ of a sphere centered at $r=0$ is  given by the product of $(L_{\rm AdS}\sinh(r/L_{\rm AdS}))^{d-2}$ times the area of a unit $S^{d-2}$. Assuming a point charge $Q$ is located at $r=0$, according to Gauss law the spherically symmetric electric field $E$ is such that $E \simeq Q/A \sim (\sinh(r/L_{\rm AdS}))^{2-d}$ and thus the electric potential $\phi_{\rm E}$ scales like $\phi_{\rm E}\simeq (\sinh(r/L_{\rm AdS}))^{3-d}$. For $r \ll L_{\rm AdS}$, we recover the usual flat space result $\phi_{\rm E} \simeq \frac{1}{r^{d-3}}$, while for $r \gg L_{\rm AdS}$ we have $\phi_{\rm E} \simeq e^{-(d-3)r}$. Therefore, at large distances in AdS the electric interaction is exponentially suppressed, while in flat space it is only power-law suppressed. 
This means that any particle, even massless, at a distance $r> L_{\rm AdS}$ is effectively not interacting electromagnetically with a charge at $r=0$. We can thus think of AdS as a box of size $L_{\rm AdS}$, or we can say that, with respect to flat space, AdS behaves as an IR regulator according to the effective prescription
\begin{equation}
r \to L_{\rm AdS}\sinh(r/L_{\rm AdS})\, .
\end{equation}

A similar conclusion on $L_{\rm AdS}$ being the IR cutoff is reached when looking at gravitational interactions. A massive particle moving in AdS feels a gravitational potential $\phi_G \simeq 1+ \rho^2/L_{\rm AdS}^2$, with $\rho = L_{\rm AdS}\sinh(r/L_{\rm AdS})$. Since this diverges at infinity, it effectively confines the particle inside a box of size $L_{\rm AdS} \simeq \Lambda_{\rm IR}^{-1}$. For massless particles one can impose reflecting boundary conditions at infinity and reach once more the same conclusion.

Are these heuristic arguments enough to conclude that $\Lambda_{\rm IR} \simeq L_{\rm AdS}^{-1}$? The existence of large AdS black holes poses a challenge. They have radius $L > L_{\rm AdS}$ and as such they would probe distances larger than $\Lambda_{\rm IR}^{-1}$. Furthermore, they are thermodynamically stable because any emitted particle is absorbed due to the confining properties of AdS just discussed  \cite{Hemming:2007yq}. Hence, there exist stable objects probing lengths larger than $L_{\rm AdS}$.
Since we are considering large black holes, they will definitely contain the physical system we are studying. This means that the extensive EFT entropy is upper bounded by the entropy of the black hole with radius $L$, 
\begin{equation} 
(\Lambda_{\rm UV}L)^{d-1} \leq N (\Lambda_{\rm UV}L)^{d-1} < (M_{\rm Pl,d}L)^{d-2}\,.
\end{equation}
In fact, we can surely demand this for the largest possible black hole. Inside its horizon there are regions in which the assumption on the EFT being weakly coupled fails. However, if the black hole is really large and if the systems is localized somewhere close to the horizon, one might hope that the gravitational interaction therein is still dominated by the AdS curvature in such a way that our starting assumption still hold. We proceed with this option in mind.  
For large black holes $L>L_{\rm AdS}$, and thus we can write
\begin{equation}
M_{\rm Pl,d}^{2-d} \Lambda_{\rm UV}^{d-1} \leq L^{-1}< L_{\rm AdS}^{-1}\, .
\end{equation}
This implies that the relation \eqref{eq:CEBd} in which $\Lambda_{\rm IR} \simeq L_{\rm AdS}^{-1}$ is possibly justified even in the presence of large AdS black holes. In fact, if $L_{\rm AdS}$ is an underestimate for the IR energy scale, then the considered theories should be satisfying the bound as strict inequality.

All of the above arguments are nevertheless heuristic. In the next section, we thus present a more convincing derivation of the bound  \eqref{eq:CEBd} which does not rely on them and which it will allow us to better understand its domain of validity.

\subsection{A derivation from ten dimensions}
\label{sec:derbound}

In this section, we provide a derivation of a bound closely related to \eqref{eq:CEBd} in string compactifications, under a few assumptions and along the lines sketched in \cite{Cribiori:2024jwq}. We work in type II string theory, but the analysis can be translated to eleven-dimensional supergravity, type I and heterotic string theory by appropriately setting to zero the appropriate fields. The key ingredient in the derivation will be fundamental domain walls; hence we will name the resulting inequality the \emph{domain wall bound}.

Let us start from the ten-dimensional equations of motion in Einstein frame (we follow the conventions of \cite{Blaback:2010sj}) 
\begin{equation}\begin{split}
    R_{MN}=\,&\frac{1}{2}\partial_M\phi\partial_N\phi+\frac{1}{2}e^{-\phi}\left(|H_3|^2_{MN}-\frac{1}{4}g_{MN}|H_3|^2\right)\\
    &+\sum_{n\leq 5} e^{\frac{5-n}{2}\phi}\left(\frac{1}{2(1+\delta_{n5})}|F_n|^2_{MN}-\frac{n-1}{16(1+\delta_{n5})}g_{MN}|F_n|^2\right)\\
    &+\frac12 \left(T_{MN}^{loc} - \frac14 g_{MN}T^{loc}\right)\,,
    \end{split}
\end{equation}
where the contribution of spacetime filling $Dp$/$Op$ sources with tension proportional to $\mu_p>0$ and wrapping an internal $(p-3)$-cycle $\Sigma$ is $T^{\rm loc}_{MN} = \mp \sum_p\mu_p\, e^{\frac{p-3}{4}\phi} \Pi_{MN}\delta(\Sigma)$, and we set $T^{\rm loc} = g^{MN}T_{MN}^{\rm loc}$. In our conventions, the upper sign is for D-branes while the lower sign for O-planes. We also have the dilaton equation
\begin{equation}
\begin{split}
\nabla^2 \phi = &-\frac12 e^{-\phi}|H_3|^2 +\sum_{n\leq 5}\left(\frac{5-n}{4(1+\delta_{n5})}\right)e^{\frac{5-n}{2}\phi}|F_n|^2 \\
&\pm \sum_{p\geq 3}  \left(\frac{p-3}{4}\right) \mu_p \,e^{\frac{p-3}{4}\phi}\delta (\Sigma)\,.
\end{split}
\end{equation}

We consider a compactification to $d<10$ dimensions and containing a $d$-dimensional external AdS factor. In order not to break maximal symmetry, all $F_n$ fluxes are taken to be internal, with possible exception for $n=d$ (and $H_3$ for $d=3$). In all cases, we trade external flux, such as $F_d$, for internal one, such as $F_{10-d}$, without explicitly writing it. With this in mind, we calculate the $d$-dimensional Ricci scalar
\begin{equation}
\begin{split}
R_d = &-\frac d8 e^{-\phi}|H_3|^2 -\frac{d}{16}\sum_n e^{\frac{5-n}{2}\phi}\frac{n-1}{1+\delta_{n5}}|F_n|^2\mp \frac{d}{16} \sum_{p\geq 3} \mu_p(7-p) e^{\frac{p-3}{4}\phi} \delta (\Sigma)\,.
\end{split}
\end{equation}
Using the dilaton equation under the assumption of constant $\phi$, 
\begin{equation}
-\frac d8 e^{-\phi}|H_3|^2 + \frac{d}{16} \sum_n e^{\frac{5-n}{2}\phi}\left(\frac{5-n}{1+\delta_{n5}}\right)|F_n|^2 \pm \frac{d}{16}\sum_{p\geq 3} \mu_p (p-3) e^{\frac{p-3}{4}\phi} \delta(\Sigma)=0\,,
\end{equation}
we eventually get
\begin{equation}
R_{d} = -\frac d4 \sum_n \left(\frac{1}{1+\delta_{n5}}\right)e^{\frac{5-n}{2}\phi}|F_n|^2 \mp \frac d4 \sum_{p\geq 3} \mu_p e^{\frac{p-3}{4}\phi} \delta (\Sigma)\,,
\end{equation}
where the minus sign is for positive tension sources, while the plus sign for negative tension. The approximation of constant dilaton is similar to the approximation of no warping. The conclusions one can draw from these assumptions are equivalent to working at the level of the equations integrated  over the compact space \cite{Blaback:2010sj}. In this sense, it is not overly restrictive. 
Recall that for an AdS geometry $R_d = -d(d-1)/L_{\rm AdS}^2$. Hence, we find
\begin{equation}
\frac{d-1}{L_{\rm AdS}^2} = \frac14  \sum_n \left(\frac{1}{1+\delta_{n5}}\right)e^{\frac{5-n}{2}\phi}|F_n|^2 \pm \frac 14 \sum_{p\geq 3} \mu_p e^{\frac{p-3}{4}\phi} \delta (\Sigma)\,.
\end{equation}

If sources are only D-branes, we just have positive terms on the right hand side (recall $\mu_p>0$) and thus write
\begin{equation}\label{Ricciscalar s.u.}
\frac{d-1}{L_{\rm AdS}^{2}} \geq  \frac14  \sum_n \left(\frac{1}{1+\delta_{n5}}\right)e^{\frac{5-n}{2}\phi}|F_n|^2\,,
\end{equation}
or, in string frame,\footnote{We pass from Einstein to string frame via 
\begin{equation}
    g_{MN}^E=e^{-\phi/2}g_{MN}^S\,,\quad |H_3^E|^2=e^{3\phi/2}|H_3^S|^2\,,\quad |F_n^E|^2=e^{n\phi/2}|F_n^S|^2\,,\quad R^E=e^{\phi/2}R^S\,,
\end{equation}
where $E$ and $S$ correspond to Einstein and string frame respectively. We omitted the superscript $E$ in the main text. Recall also that the tension of a $Dp$-brane in Einstein frame is $T_p\simeq M_s^{p+1}g_s^{\frac{p-3}{4}}$.}
\begin{equation}
    \frac{d-1}{L_{\rm AdS}^{2}} \geq  \frac14  \sum_n \left(\frac{1}{1+\delta_{n5}}\right)e^{{2}\phi}|F_n^S|^2\,.
\end{equation}
If sources include O-planes  we also have negative terms and cancellations can occur, hence we have to be more careful. 
To have an AdS solution, the flux terms should not be exactly canceled (nor be dominated) by the O-plane contribution. This generically happens if all terms are of the same order, as in a form of detailed balance, and thus the on-shell value of $1/L_{\rm AdS}^2$ is proportional to $|F_n|^2$ up to some model-dependent coefficient $c_n>0$. Classical solutions of the ten-dimensional equations of motion can satisfy this property. A well-known exception are no-scale models, which we will comment on later.  
Barring fine tuning, we thus write
\begin{equation}
\label{eq:LadsFn2}
\frac{1}{L_{\rm AdS}^2} \geq  c_{n} \, e^{2\phi}|F_{n}^S|^2 \,, \qquad c_n = \mathcal{O}(1)\,.
\end{equation}

To have a consistent solution of string theory we have to impose a quantization condition on the magnetic flux $F_n$,
\begin{equation}
    \int_{X_n}F_n^S=M_s^{1-n}f_n\,,
\end{equation}
where $M_s$  is the string mass and $f_n$ is an integer. This condition implies that
\begin{equation}
    |F_n^S|^2=\left( \frac{M_s^{1-n}f_n}{{\rm Vol}(X_n)}\right)^2 = M_s^2 \left(\frac{f_n}{{\rm vol}(X_n)}\right)^2\,,
\end{equation}
where $X_n$ is the $n$-cycle threaded by the flux. We indicate with $\text{Vol}(X_n)$ the dimensionful volume of $X_n$ and with $\text{vol}(X_n)$ the volume in string units. Recalling that $M_{\rm Pl,d}^{d-2} = g_s^{-2}\text{vol}_{10-d} M_s^{d-2}$ with $\text{vol}_{10-d}$ the total compact volume in string units, we can rewrite \eqref{eq:LadsFn2} as
\begin{equation}
\begin{aligned}
 \frac{M_{\rm Pl,d}^{2(d-2)}}{L_{\rm AdS}^2} &\geq c_n\, M_s^{2(d-1)}f_n^2 \,\,e^{-2\phi}\left(\frac{{\rm vol}_{10-d}}{{\rm vol}(X_n)}\right)^2  \\
 &=c_n\, M_s^{2(d-1)}f_n^2 \,\,e^{-2\phi}\left(\frac{{\rm vol}_{10-d}}{{\rm vol}(Y_p)}\right)^2\left(\frac{{\rm vol}(Y_p)}{{\rm vol}(X_n)}\right)^2\\
 &=c_n\, f_n^2\, \left(\frac{{\rm vol}_{10-d}}{{\rm vol}(Y_p)}\right)^2 T_{\rm dw}^2\,.
 \end{aligned}
\end{equation}
In the last line, we identified the string frame tension, $T_{\rm dw}$, of a domain wall in $d$-dimensions obtained by wrapping a $D(8-n)$-brane on a $p$-cycle $Y_p$ and smearing it on an 
$n$-cycle $X_n$, with $8-n = d+p-2$,
\begin{equation}
T_{\rm dw} = \frac{M_s^{d-1}}{e^\phi} \frac{{\rm vol}(Y_p)}{{\rm vol}(X_n)}\, .
\end{equation}
Notice that at this step we assume $d>2$ to avoid subtleties with the definition of $M_{\rm Pl,2}$; the two-dimensional case has been considered already in \cite{Cribiori:2024jwq}. 
Since $Y_p$ is a cycle internal to the compact manifold, we surely have ${\rm vol}_{10-d}> {\rm vol}(Y_p) \gg 1$, where the last inequality is needed for validity of the supergravity approximation and is valid for a product manifold. Hence, since $f_n$ is integer and $c_n = \mathcal{O}(1)$,  we arrive at
\begin{equation}
 \frac{M_{\rm Pl,d}^{2(d-2)}}{L_{\rm AdS}^2}  \geq T_{\rm dw}^2\, .
\end{equation}

Assuming that the domain wall is fundamental, namely that it cannot be resolved by the $d$-dimensional EFT, its tension sets an upper bound for the UV cutoff, $T_{\rm dw}\geq \Lambda_{\rm UV}^{d-1}$, giving
\begin{equation}
 \frac{M_{\rm Pl,d}^{d-2}}{L_{\rm AdS}}  \geq \Lambda_{\rm UV}^{d-1}\,.
\end{equation}
This is the anticipated \emph{domain wall bound} and it matches with \eqref{eq:CEBd} upon the identification $\Lambda_{\rm IR} \simeq 1/L_{\rm AdS}$. 
Hence, in view of our discussion on the AdS radius playing the role of the IR cutoff, we find that \eqref{eq:CEBd} is automatically satisfied for AdS flux vacua of string theory in which the detailed balance condition holds and which are interpolated by fundamental BPS domain walls. For $d=2$, the above relation forbids scale separation, as noted in \cite{Cribiori:2024jwq}.

We now comment on the limitations of our analysis. 
The way we introduced a domain wall tension into the derivation relies on the BPS condition relating charge and tension to one another. For non-supersymmetric vacua, such as the Large Volume Scenario AdS vacuum \cite{Balasubramanian:2005zx}, we can instead exploit the weak gravity conjecture extended to domain walls \cite{Ooguri:2016pdq}. It holds that a non-supersymmetric domain wall is characterized by a tension $T_{\rm dw}$ upper bounded by that of a BPS one,
\begin{equation}
T_{\rm dw} \leq \frac{M_s^{d-1}}{e^\phi} \frac{{\rm vol}(Y_p)}{{\rm vol}(X_n)}\, ,
\end{equation}
which again leads to the same domain wall bound.\footnote{This form of the weak gravity conjecture  has been contested for instance in \cite{Narayan:2010em, Giombi:2017mxl, Casas:2022mnz, Montero:2024qtz, Menet:2025nbf}, but it is not clear whether all effects that can contribute to domain wall tensions have been taken into account \cite{Marchesano:2022rpr}; this is the case especially for polarization effects which can lead to new domain wall decays \cite{Menet:2025mbi}.}

Our derivation holds most efficiently for vacua with only D-branes, while in the presence of O-planes we additionally assumed a detailed balance condition, namely $c_n = \mathcal{O}(1) >0$. Solutions with negative tension objects and featuring $c_n=0$ are a trivial case that lead to no paradox since anyway $1/L_{\rm AdS}^2 \geq 0$, with the inequality being saturated for Minkowski. Well-known examples of this kind are no-scale solutions, such as \cite{Giddings:2001yu, Blaback:2010sj, Farakos:2020phe, Grana:2006kf}. However, there is no reason to believe that these are solutions of string theory, given that the no-scale structure in the absence of extended supersymmetry hardly survives even perturbative corrections. One might thus wonder about the situation in which $c_n=0$ classically, but an AdS vacuum is produced after corrections are included. This is the case in the celebrated KKLT scenario \cite{Kachru:2003aw}, but also in the broader class of racetrack models \cite{Blanco-Pillado:2004aap,Kallosh:2004yh} or the LVS construction \cite{Balasubramanian:2005zx}.  We believe that our bound might be reliable also in these examples for various reasons. At its core, our derivation exploits the existence of a domain wall interpolating between vacua. We are not aware of any obstruction to the existence of a such an object in KKLT or racetrack models. For example, in type IIB compactifications (KKLT, LVS) the domain wall discharges magnetic $H_3$ and $F_3$ flux and is thus a bound state of $NS5$- and $D5$-branes \cite{Gukov:1999ya}.
Even if these vacua feature an interplay between classical terms and quantum corrections, it is possible that the dynamics links the two. This seems to be the case in the simplest KKLT construction but also in more recent refinements, as we will discuss in section \ref{sec:boundKKLT}.

\subsection{A lower bound on the gravitino mass} 
\label{eq:boundm32}

An important consequence of the domain wall bound derived in the previous section,
\begin{equation}
\label{eq:DWBLads}
\Lambda_{\rm UV}^{d-1} \leq M_{\rm Pl,d}^{d-2}L_{\rm AdS}^{-1}\,,
\end{equation}
is that it provides a lower bound on the gravitino mass parameter in supersymmetric AdS vacua, $1/{L_{\rm AdS}^2}\simeq m_{3/2}^2$, namely
\begin{equation}
\label{eq:CEBm32}
\Lambda_{\rm UV}^{d-1}/M_{\rm Pl, d}^{d-2} \leq m_{3/2}\,.
\end{equation}
This implies that the limit $m_{3/2} \to 0$ at fixed $M_{\rm Pl,d}$ must be accompanied by $\Lambda_{\rm UV} \to 0$, leading to the breakdown of the EFT description. This is in line with previous investigations saying that an arbitrarily small gravitino mass is problematic for the EFT \cite{Cribiori:2020use,DallAgata:2021nnr}. It is also the physical content of the gravitino conjecture \cite{Cribiori:2021gbf,Castellano:2021yye} (and of the anti-de Sitter distance conjecture \cite{Lust:2019zwm}, when they overlap), of which we have thus provided a derivation from string compactifications, under certain assumptions.

Specializing to minimal supergravity in four dimensions with K\"ahler potential $K$ and superpotential $W$, we have $M_{\rm Pl,4}^{4}\,m_{3/2}^2= e^{K/M_{\rm Pl,4}^2}|W|^2$, and thus we can rewrite \eqref{eq:CEBm32} as\footnote{In our conventions $W$ has mass dimension 3 and $K$ has mass dimension 2.}
\begin{equation}
\Lambda_{\rm UV}^{3} <e^{\frac12 K/{M_{{\rm Pl},4}^2}} |W|\,.
\end{equation}
A well-studied example is the Gukov-Vafa-Witten superpotential, arising in compactifications of type II string theory on manifolds with SU(3) structure and $G_3=F_3 -i e^{-\phi}H_3$ flux,
\begin{equation}
W_{0}= M_{{\rm Pl},4}^8\int \Omega \wedge G_3\,,
\end{equation}
where $\Omega$ is the unique (up to rescaling) holomorphic $(3,0)$-form on the compact manifold. Therefore, in this case the domain wall bound translates into a lower bound on $e^{\frac12 K/{M_{\rm Pl,4}^2}}|W_{0}|$. We will discuss in section \ref{sec:boundKKLT} the consequences of this bound when trying to realize the KKLT scenario. 

\subsection{Counting AdS vacua}

The picture we have in mind is a collection of $d$-dimensional AdS vacua interpolated by domain walls. This is motivated by the heuristic idea that all vacua in quantum gravity should be connected to one another by some process or mechanism; in our case, this would be a domain wall transition. 

For simplicity, we assume that each domain wall discharges one unit of flux such that $f_n$ can be used as a proxy for the number of domain wall transitions. Then, the domain wall bound \eqref{eq:DWBLads} provides us with a counting of the number of AdS vacua (or, say, of domain walls) up to the UV cutoff of the $d$-dimensional EFT. From the derivation in section \ref{sec:derbound}, we have 
\begin{equation}
    f_n^2\leq \frac{M_{{\rm Pl},d}^{2(d-2)}}{\Lambda_{\rm UV}^{2(d-1)}}\Lambda_{\rm IR}^2< \frac{M_{{\rm Pl},d}^{2(d-2)}}{\Lambda_{\rm UV}^{2(d-2)}}\,,
\end{equation}
where in the last step we used the EFT condition $\Lambda_{\rm IR}<\Lambda_{\rm UV}$. If we identify the UV cutoff with the species scale and use the formula for the latter given in \cite{Veneziano:2001ah, Dvali:2007hz}, namely $\Lambda_{\rm sp}=M_{\rm Pl,d}/N_{\rm sp}^{\frac{1}{d-2}}$, we find that the number of light species upper bounds the number of fluxes counting the AdS vacua, 
\begin{equation}
    f_n<N_{\rm sp}\,.
\end{equation}
However, since the formula of \cite{Veneziano:2001ah, Dvali:2007hz} is quite heuristic, it is perhaps better to just estimate the number of vacua in terms of the (model-dependent) UV cutoff; similar ideas appeared $e.g.$~in \cite{Acharya:2006zw,Delgado:2024skw,Baykara:2025nnc}. To this purpose, we introduce the dimensionless quantity $\lambda_{\rm UV}=\frac{\Lambda_{\rm UV}}{M_{{\rm Pl},d}}\leq 1$ and thus obtain 
\begin{equation}
\text{\# of AdS vacua}  \sim   f_n< \lambda_{\rm UV}^{-(d-2)}\, .
\end{equation}
Hence, the domain wall bound \eqref{eq:DWBLads} tells us that the number of AdS vacua is bounded by a polynomial in the UV cutoff whose exponent depends only on the number of spacetime dimensions in which the EFT is defined (recall that in this work we assume $d>2$).

\section{The bound on classical flux vacua}

As first application, we check if the domain wall bound \eqref{eq:DWBLads} is satisfied in classical flux vacua with scale separation, namely with a hierarchy between the Kaluza-Klein scale, $L_{\rm KK}=1/m_{\rm KK}$, and $L_{\rm AdS}$. 
In four dimensions, one has the so-called DGKT vacua \cite{DeWolfe:2005uu, Camara:2005dc}, also in their non-isotropic version \cite{Carrasco:2023hta,Tringas:2023vzn}, and the more recent class in massless type IIA and eleven-dimensional supergravity \cite{Cribiori:2021djm}.\footnote{An analysis of the covariant entropy bound in DGKT was performed in \cite{Castellano:2021mmx}; here we go beyond it also by considering anisotropies \cite{Carrasco:2023hta,Tringas:2023vzn} and the class of models in massless IIA supergravity \cite{Cribiori:2021djm,VanHemelryck:2024bas}.} In three dimensions, we look at the vacua of \cite{Farakos:2020phe} (see also \cite{VanHemelryck:2022ynr}), but it would be interesting to extend the discussion to the very recent constructions \cite{VanHemelryck:2025qok,Miao:2025rgf,Tringas:2025bwe, Arboleya:2024vnp, Arboleya:2025ocb, Arboleya:2025jko, Proust:2025vmv, Farakos:2025bwf}. The result of this analysis is to be expected: all of these vacua satisfy a detailed balance condition and thus they will also satisfy the domain wall bound \eqref{eq:DWBLads}, but possibly without saturating it. It is nevertheless interesting to see how the various scales conspire towards the result, and to discuss different possible microscopic realizations of the interpolating domain wall. If we think of the domain wall bound as motivated by holography, via $e.g.$ the covariant entropy bound, it is it noteworthy that all of these models satisfy it, even if none of them have a known holographic dual \cite{Aharony:2008wz,Conlon:2021cjk,Apers:2022zjx,Bobev:2023dwx,Bobev:2025yxp}.

\subsection{Domain wall tensions}

The DGKT construction \cite{DeWolfe:2005uu, Camara:2005dc} is a compactification of massive type IIA supergravity on a Calabi-Yau orientifold $X_6$. It gives rise to a one-parameter family of four-dimensional AdS vacua labelled by an unconstrained integer $N$,
\begin{equation}
N\simeq \int_{X_4} F_4^S\,.
\end{equation}
We will be working in string units and the limit $N \to \infty$ is understood in what follows.  
We first consider the isotropic setup in which all 2-cycle volumes, ${\rm vol}(X_2)\equiv v$, scale with the same power, $v \sim N^\frac12$. The physical scales and the string coupling then behave as
\begin{equation}
\label{eq:DGKTscaling}
L_{\rm KK} \sim N^{\frac14}\,, \qquad L_{\rm AdS} \sim N^{\frac34}\,, \qquad g_s \sim N^{-\frac34}\,, \qquad M_{\rm Pl,4}^2 \simeq g_s^{-2} L_{\rm KK}^6  \sim N^3 \,,
\end{equation}
with ${\rm Vol} (X_6) \simeq L_{\rm KK}^6$ and $g_s = e^\phi$. 

The vacuum energy is sustained by magnetic $F_4$, $F_0$ or $H_3$ flux whose strength is balanced on-shell. Thus, in principle one can consider domain walls obtained from wrapping $D4$-, $D8$- or $NS5$-branes on $2$-, $6$- or $3$-cycles respectively. 
Here, we will only look at the naive scaling of the tension assuming that these domain walls are BPS; the whole story is much more subtle and we refer to \cite{Marchesano:2021ycx,Casas:2022mnz,Marchesano:2022rpr} for a more complete and precise discussion. 
In string units the corresponding domain wall tensions are
\begin{equation}
T_{D4,{\rm dw}} = L_{\rm KK}^2/g_s\sim N^{5/4}\,,\quad T_{D8,{\rm dw}}=L_{\rm KK}^6/g_s\sim N^{9/4}\,,\quad T_{NS5,{\rm dw}}=L_{\rm KK}^3/g_s^2  \sim N^{9/4}\,.   
\end{equation}
The $D4$-brane domain walls are the lightest, with $NT_{D4,{\rm dw}}\sim T_{D8,\rm{dw}}\sim T_{NS5,\rm{dw}}\sim (N^{3/4})^3$. 
The third root of their tension sets an upper bound on the UV scale, 
\begin{equation}
\Lambda_{\rm UV}<N^{5/12}<N^{3/4}\,,    
\end{equation}
which we will check in the following section. 

As of today, the most explicit realization of the DGKT construction is given by toroidal orbifolds $X_6 = T^6/\Gamma$ with $\Gamma \subset SU(3)$. Besides the isotropic setup discussed above, scale separation has been found also in the non-isotropic configuration in which the volume of each of the three 2-tori scales with a different power of $N$ \cite{Carrasco:2023hta, Tringas:2023vzn}. Denoting again these volumes with $v_i$, $i=1,2,3$, we have three real scaling parameters $f_i$ such that
\begin{equation}\label{eq:DGKTnonisotr1}
v_1 \sim N^{\frac12(-f_1+f_2+f_3)}\,, \qquad v_2 \sim N^{\frac12(f_1-f_2+f_3)}\,, \qquad v_3 \sim N^{\frac12(f_1+f_2-f_3)}
\end{equation}
and also
\begin{equation}\label{eq:DGKTnonisotr2}
L_{{\rm KK},i}\sim v_i^{\frac12}\,, \qquad g_s \sim N^{-\frac14 (f_1+f_2+f_3)}\,, \qquad L_{\rm AdS} \sim N^{\frac14 (f_1+f_2+f_3)}\,.
\end{equation}
Since $L_{{\rm KK},i}/L_{\rm AdS}\sim N^{-\frac{f_i}{2}}$ scale separation is achieved for $f_i>0$, while $\sum_i f_i>0$ gives large volume and small string coupling. The isotropic setup is $f_1=f_2=f_3=1$.  The scalings of the $D4$, $D8$ and $NS5$ domain walls are modified analogously.

Next, we consider the class of four-dimensional vacua arising from compactification of massless IIA supergravity on the Iwasawa manifold, which is a specific example of twisted torus \cite{Kounnas:2007dd,Caviezel:2008ik,Cribiori:2021djm}. To achieve scale separation, one is here forced to consider non-isotropic fluxes from the very beginning. Remarkably, solutions at both weak and strong string coupling are possible \cite{Cribiori:2021djm}. Denoting the volumes of the 2-tori in strings units with $v_i,i=1,2,3$,  and introducing real parameters $f_i$, the scalings are
\begin{equation}\label{eq:DGKTIa}
v_1 \sim N^{\frac{1}{2}(f_1-f_2-f_3)}\,, \qquad v_2 \sim N^{\frac{1}{2}(f_1-f_2+f_3)}\,, \qquad v_3 \sim N^{\frac{1}{2}(f_1+f_2-f_3)}\,,
\end{equation}
and also
\begin{equation}\label{eq:DGKTIb}
L_{\rm KK} \sim v_2^\frac12\,, \qquad g_s \sim N^{\frac{1}{4}(f_1-3f_2-3f_3)}\,, \qquad L_{\rm AdS} \sim N^{\frac14 (f_1+f_2+f_3)}\,,
\end{equation}
where $f_i\geq 0$ and without loss of generality $f_3 \geq f_2 $ ($N$ is here such that $N^{f_1}\simeq \int_{X_6}F_6$).  Possible domain walls in this model are $D2$-  and $D6$-branes wrapping 4-cycles \cite{Caviezel:2008ik,Apers:2022vfp} with scalings 
\begin{equation}
    T_{D2, {\rm dw}}= 1/g_s\sim N^{\frac{1}{4}(3f_2+3f_3-f_1)}\,,\quad T_{D6,{\rm dw},1i}= v_1v_i/g_s\sim N^{\frac{1}{4}(3f_j+3f_k-f_i)}\,,
\end{equation}
so that we have a bound on the UV cutoff
\begin{equation}
    \Lambda_{\rm UV}<N^{\frac{1}{12}(3f_2+3f_3-f_1)}\,,
\end{equation}
which we will verify in the following section.

We have just confirmed the lesson of \cite{Caviezel:2008ik,Apers:2022vfp}: replacing 4-form fluxes with intersecting $D4$-branes and then taking a near horizon limit leads to spacetimes whose scaling of the dilaton, volume moduli and AdS length scale is as in DGKT. An analogous story holds for the compactification on the Iwasawa manifold. 
This is somewhat confusing, since there is no scale-separated vacuum without $O6$-planes and the extra ingredients such as the Romans mass and $H_3$ flux for DGKT, or the non-trivial second torsion class for the Iwasawa manifold. 
However, this lesson seems to suggest that changing to branes the fluxes whose large $N$ limit is considered can correctly predict the scalings on the AdS vacuum.

\subsection{Four- and three-dimensional vacua}

Having discussed possible domain walls in the previous section, we proceed now by checking  the domain wall bound \eqref{eq:DWBLads} on various vacua. 
We identify $\Lambda_{\rm UV}$ with the species scale \cite{Veneziano:2001ah,Dvali:2007hz}, which gives an upper bound on the UV cutoff of gravitational effective theories. If the domain wall bound is violated for the species scale, it is certainly violated for any other UV cutoff that is necessarily lower.

We start from the most studied DGKT construction, namely the original isotropic setup with scalings \eqref{eq:DGKTscaling}. Hence, we have to determine the species scale $\Lambda_{\rm UV}$. There are two main candidates: the string scale $M_s$ and  ten-dimensional Planck scale $M_{\rm Pl, 10}$. Given that $M_s \sim 1$, while  $M_{\rm Pl, 10}\sim g_{s}^{-\frac14}\sim N^{\frac{3}{16}}\gg1$, the string scale is lower than the ten-dimensional Planck scale and thus it contributes a large number of states to the species scale. Since these states are string oscillators, whose degeneracy is exponential, they will surely dominate any possible Kaluza-Klein tower whose degeneracy is polynomial. We conclude that the species scale in this setup is the string scale, $\Lambda_{\rm UV} = M_s \sim 1$. The domain wall bound \eqref{eq:DWBLads} reduces then to
\begin{equation}
1 \leq g_s^{-2} L_{\rm KK}^6/L_{\rm AdS}
\end{equation}
and one can check that it is indeed satisfied for the scalings \eqref{eq:DGKTscaling}. 

Considering non-isotropic fluxes and using \eqref{eq:DGKTnonisotr1} and \eqref{eq:DGKTnonisotr2}, we reach a similar conclusion. Namely, the ten-dimensional Planck scale $M_{\rm Pl,10} \sim N^{\frac{1}{16}(f_1+f_2+f_3)}\gg1$ is heavier than the string scale $M_s \sim 1$ and thus the species scale is the latter. Since the four-dimensional Planck mass is ${M^2_{\rm Pl,4}}\sim N^{f_1+f_2+f_3}$, the domain wall bound \eqref{eq:DWBLads} is satisfied in the regime of validity of the supergravity approximation.

Next, we consider the  four-dimensional vacua arising from compactification of massless IIA supergravity on the Iwasawa manifold described by \eqref{eq:DGKTIa} and \eqref{eq:DGKTIb}. In this setup, the four-dimensional Planck mass is $M_{\rm Pl,4}^2 = g_s^{-2}{\rm vol}_6 \sim N^{f_1+f_2+f_3}$. The strong coupling solution corresponds to $f_1 > 3(f_2+f_3)$ and in this case the species scale is the eleven-dimensional Planck scale $M_{\rm Pl, 11} \sim g_s^{-\frac13} < 1$ which is indeed smaller than the string scale $M_s \sim 1$. The domain wall bound  \eqref{eq:DWBLads} reduces to $1\leq g_s M_{\rm Pl,4}^2/L_{\rm AdS}$ and one can check that it is satisfied.
The weak coupling solution corresponds to $f_1 < 3(f_2+f_3)$ and thus $M_{\rm Pl, 11}>M_s \sim 1$ and here the species scale is the string scale. The domain wall bound  \eqref{eq:DWBLads} reduces to $1\leq M_{\rm Pl,4}^2/L_{\rm AdS}$ and one can check that it is satisfied. 

Finally, we look at the three-dimensional vacua of \cite{Farakos:2020phe}. They have the same scalings as the isotropic DGKT model
\begin{equation}
g_s\sim N^{-3/4}\,,\qquad L_{\rm KK}\sim N^{1/4}\,,\qquad L_{\rm AdS}\sim N^{3/4}\,
\end{equation}
and thus the species scale is again the string scale, which is smaller than the ten-dimensional Planck mass $M_{\rm Pl,10}\sim g_s^{-\frac14} \sim N^{\frac{3}{16}}$. Since $M_{\rm Pl, 3} = g_s^{-2}L_{\rm KK}^7$, the domain wall bound \eqref{eq:DWBLads} reduces to
\begin{equation}
1 \leq g_s^{-2} L_{\rm KK}^7/L_{\rm AdS}
\end{equation}
and one can check that it is satisfied.

\section{The bound beyond classical flux vacua}
\label{sec:boundKKLT}

The domain wall bound \eqref{eq:DWBLads} represents a challenge if one is to engineer very large $L_{\rm AdS} M_{\rm Pl, 4} $ while at the same time $\Lambda_{\rm UV}$ does not vary appropriately. This leads us to consider KKLT-like scenarios, where one aims at achieving a very small value (in units of $M_{\rm Pl, 4}$) of the classical contribution to the gravitino mass, while the UV cutoff typically depends very loosely on it. This leads to a violation of the lower bound on the gravitino mass presented in section \ref{eq:boundm32}.

The KKLT scenario \cite{Kachru:2003aw} is a proposal for moduli stabilization in string compactifications eventually leading to a de Sitter minimum. Concretely, it is settled in Calabi-Yau orientifolds compactifications to four dimensions of weakly coupled type IIB string theory in the presence of fluxes and (anti-)D-branes. Here, one engineers a supersymmetric, four-dimensional AdS vacuum with small cosmological constant in units of $M_{\rm Pl, 4}$. We will first consider the simplest, although abstract, models and then proceed with more sophisticated and explicit constructions, in particular \cite{Demirtas:2019sip,Demirtas:2020ffz,Alvarez-Garcia:2020pxd,Demirtas:2021nlu,Demirtas:2021ote,McAllister:2024lnt}. 
We will therefore distinguish between KKLT-like AdS vacua with and without a strongly warped throat in the Calabi-Yau geometry. In this section, we work in units of $M_{\rm Pl,4}$.

\subsection*{Large Volume Scenario}

Before entering the details of KKLT, we comment briefly on a popular alternative, namely the Large Volume Scenario (LVS) \cite{Balasubramanian:2005zx}. The starting point is a no-scale compactification with 3-form fluxes on Calabi-Yau orientifolds in weakly coupled type IIB string theory; then, moduli are stabilized on a non-supersymmetric AdS vacuum. 
In LVS, finetuning and control over the scenario stem from inverse volume suppression. We will skip any detail and just check the naive consistency with the domain wall bound, which superficially seems to be satisfied. Working in units of $M_{\rm Pl,4}$ and denoting $\Lambda \simeq 1/L_{\rm AdS}^2$ the AdS cosmological constant, the model features the Kaluza-Klein scale $m_{\rm KK}\simeq \Lambda^{\frac 29}$ (up to logarithmic corrections) \cite{Blumenhagen:2019vgj} and the ten-dimensional Planck scale $M_{\rm Pl,10}\simeq m_{\rm KK}^{\frac{3}{4}}\simeq \Lambda^{\frac16}$. Taking (erroneously) $\Lambda_{\rm UV}\simeq M_{\rm Pl, 10}$, one can check that the domain wall bound \eqref{eq:DWBLads} is saturated. This implies that the bound is safe for the appropriate cutoff, which is the species scale $\Lambda_{\rm sp}\simeq M_s< M_{\rm Pl,10}$.

\subsection{Basic KKLT model}

KKLT-like AdS vacua arise from balancing two terms, a classical flux contribution and appropriate non-perturbative quantum corrections. If we neglect the latter, the classical piece is associated to a no-scale structure \cite{Giddings:2001yu} such that, if one repeats our derivation in section \ref{sec:derbound}, one finds $c_n=0$. 
We will argue that the aforementioned quantum corrections contribute to the coefficient $c_n$ in such a way to make it non-vanishing on-shell. Key to this mechanism is really moduli stabilization of the classically flat directions. This argument motivates applying the domain wall bound also to KKLT-like AdS vacua.

We start from the prototype KKLT toy model, which is specified by the single-field K\"ahler and superpotential
\begin{equation}
K=-3\log(T+\bar T)\,, \qquad W= W_{0} + A e^{-2\pi  a  T}\,,
\end{equation}
where ${\rm Re}\,T^{3/2}$ is the volume of the Calabi-Yau orientifold, $W_{0} = \int  \Omega \wedge G_3$ the classical GVW superpotential and the exponential term in $T$ is the quantum correction. $A,a$ are real parameters of the model to be determined from the details of the microscopic theory.

By definition, $W_0$ contributes the tension of a $D5/NS5$-domain wall via the classical gravitino mass \cite{Gukov:1999ya}
\begin{equation}
T_{\rm dw}^2 = e^K W_0 \bar W_0\,.
\end{equation}
Due to the non-perturbative terms in $T$, the supersymmetric AdS vacuum energy is not exclusively flux-supported, $i.e.$ proportional to $e^K|W_0|^2$, but instead
\begin{equation}
V_{\rm AdS} = -3 e^K |W|^2= -\frac{2}{3\,{\rm Re \,}T} \pi^2 a^2 \,A^2e^{-4\pi a \,{\rm Re} T}\,.
\end{equation}
However, moduli stabilization via the F-term condition $D_T W=0$ relates the non-per\-tur\-bative terms in $T$ to the classical superpotential $W_0$ by
\begin{equation}
\label{eq:DW=0TKKLT}
-W_0 =\frac13 A e^{-2 \pi a T}(3+4\pi a \, {\rm Re}T)\,.
\end{equation}
The regime of validity of the four-dimensional EFT requires large ${\rm Re }\,T$ in string units. Since $M_{\rm Pl,4}>M_{\rm Pl, 10}>M_s$, this implies large ${\rm Re }\,T$ in Planck units. Therefore, in this regime we can approximate
\begin{equation}
-W_0 \gtrsim \frac43 \pi a \,{\rm Re} T \, A e^{-2 \pi a T}\,,
\end{equation}
and, recalling that  $e^K=1/(2\,{\rm Re} T)^3$, we then recast the tension of the $D5/NS5$-domain wall as
\begin{equation}
T_{\rm dw}^2 \simeq \frac{1}{(2\,{\rm Re }T)^3} \times \frac{16}{9} \pi^2a^2 ({\rm Re}T)^2 A^2 e^{-4\pi a t} \simeq |V_{\rm AdS}|\,.
\end{equation}
Hence, in this toy model of the KKLT scenario the vacuum energy is related to the tension of a BPS domain wall.\footnote{The existence of such domain wall requires the presence of special Lagrangian 3-cycles or of appropriate worldvolume fluxes \cite{Evslin:2007ti}.} By demanding that the domain wall is fundamental, we have
\begin{equation}
1/L_{\rm AdS}^2 \simeq T_{\rm dw}^2 \gtrsim \Lambda_{\rm UV}^6
\end{equation}
and we recover once more the domain wall bound \eqref{eq:DWBLads}.

Notice that the above argument does not really exploit the fact that $W_0$ is to be small whenever ${\rm Re}\,T$ is large. With this additional information the argument can be simplified even further. The solution \eqref{eq:DW=0TKKLT} at leading order in a small $W_0$ combined with a  large ${\rm Re} \,T$ expansion is
\begin{equation}
2\pi a \,T = - \log W_0+\dots\,.
\end{equation}
From this, we immediately find that 
\begin{equation}
\label{eq:Vads=Tdw}
|V_{\rm AdS}| = 3 e^K |W|^2 \simeq 3 e^K |W_0|^2 \simeq T_{\rm dw}^2
\end{equation}
and the domain wall bound \eqref{eq:DWBLads} follows.

The argument can be generalized to multifield. We consider the K\"ahler and superpotential 
\begin{equation}
\label{eq:EFTKKLTKaehlersec}
K = -2\log \mathcal{V}\,, \qquad W = W_0 + \sum_i A_i e^{-2\pi a_i T^i}\,,
\end{equation}
where $\mathcal{V} = \frac{1}{3!}\kappa_{ijk}t^i t^j t^k$, with $a_i$, $A_i$ and $\kappa_{ijk}$ real parameters associated to the microscopic model. Here, the $t^i$ are related to the real parts of the chiral multiplets $T_i$ as ${\rm Re}T_i = \frac{1}{2g_s}\kappa_{ijk}t^jt^k$ such that $\mathcal{V}=\frac{g_s}{3}t^i \, {\rm Re}T_i$. In hindsight, we have already introduced the string coupling $g_s$, which can be thought of as an additional parameter in this sector of the EFT. The above is sometimes termed the leading order EFT of the KKLT K\"ahler moduli. 
The conditions $D_{T_i} W=0$ are now solved by
\begin{equation}
2\pi a_i T_i = -\log W_0 +\dots\,, \qquad \forall i
\end{equation}
and therefore one can reach again the relation \eqref{eq:Vads=Tdw} and the domain wall bound \eqref{eq:DWBLads} follows.

To summarize, we showed that there is a fundamental BPS domain wall interpolating between two supersymmetric KKLT-like AdS vacua.  Hence, we expect that a consistent EFT of the KKLT scenario should satisfy the domain wall bound \eqref{eq:DWBLads}, but we will see that in some of the most explicit models available today the bound is violated. 
Notice that KKLT-like EFTs have already been argued to be non-generic in \cite{Robbins:2004hx} (especially those with a single K\"ahler modulus) or even problematic \cite{Lust:2022lfc, Blumenhagen:2022dbo,Bena:2024are,Cribiori:2025oek}.  Our analysis in the following sections will be pointing in a similar direction. In particular, \cite{Cribiori:2025oek} recently applied a holographic bound closely related to \eqref{eq:DWBLads} to the de Sitter phase of KKLT and argued that it is generically violated. Our results for the AdS phase  will be similar.

\subsection{Racetrack models and exponentially small gravitino mass}

The so far most explicit KKLT construction has been developed starting from \cite{Demirtas:2019sip} and then continuing with \cite{Demirtas:2021ote,McAllister:2024lnt}. The core idea employs twice a racetrack mechanism, once in the complex structure sector to stabilize the dilaton on a supersymmetric Minkowski vacuum, and a second time in the K\"ahler sector to stabilize the associated moduli on a supersymmetric AdS vacuum. Here, the domain wall bound \eqref{eq:DWBLads} applies, as we have argued in the previous section. Since the racetrack mechanism is the key ingredient underlying these constructions, we first discuss the restrictions posed on it by the domain wall bound abstractly and then in application to \cite{Demirtas:2021ote}. We will comment on the more recent development \cite{McAllister:2024lnt} in the next section, since it introduces a strongly warped throat as new ingredient.

Racetrack models \cite{Blanco-Pillado:2004aap,Kallosh:2004yh} are described by four-dimensional minimal supergravity with superpotential 
\begin{equation}
\label{eq:Wracetrack}
W = W_0 + Ae^{2\pi i a \tau} + Be^{2 \pi i b \tau}
\end{equation}
and K\"ahler potential $K=-\log (-i(\tau -\bar \tau))$.
Here, $W_0$ parametrizes a classical contribution, while the exponential terms are quantum corrections. These models admit supersymmetric AdS minima at $\tau = i \mathcal{V}^{2/3}$, $\mathcal{V} >0$, such that 
\begin{equation}
-W_0 = Ae^{-2\pi  a \mathcal{V}^{2/3}}(1+4\pi a \mathcal{V}^{2/3}) + Be^{-2\pi  b \mathcal{V}^{2/3}}(1+4\pi b \mathcal{V}^{2/3})
\end{equation}
giving 
\begin{equation}
\frac{1}{L_{\rm AdS}^2 } = 24 \pi^2 \mathcal{V}^{2/3} \left(a Ae^{-2\pi a \mathcal{V}^{2/3}} + bBe^{-2\pi b \mathcal{V}^{2/3}}\right)^2\,.
\end{equation}

In this section, we assume the lowest possible UV cutoff namely the Kaluza-Klein scale, $\Lambda_{\rm UV} = m_{\rm KK}$. This is a natural choice in supergravity and moreover ensures that, whenever a violation is found, any other choice of $\Lambda_{\rm UV}$, which will be necessarily higher, will lead to a violation as well. Proceeding in purely abstract terms, the relation \eqref{eq:DWBLads} translates into a bound on the parameters
\begin{equation}
m_{\rm KK}^6 \leq  24 \pi^2  \mathcal{V}^{2/3} \left(a Ae^{-2\pi a  \mathcal{V}^{2/3}} + bBe^{-2\pi b  \mathcal{V}^{2/3}}\right)^2\, ,
\end{equation}
that is telling us that $a,A,b,B$ cannot be arbitrarily chosen. Therefore, not any racetrack is automatically compatible with the domain wall bound.

The construction of AdS vacua with exponentially small gravitino mass proposed in \cite{Demirtas:2021ote,Demirtas:2021nlu} involves essentially two steps. First, one stabilizes the dilaton in a so called perturbatively flat vacuum \cite{Demirtas:2019sip}. This step involves only the complex structure sector, so that K\"ahler moduli remain massless and the vacuum is supersymmetric and Minkowski. In brief, one chooses fluxes such that the polynomial contribution to the GVW superpotential is exactly vanishing and $W_0$ is fully sustained by non-perturbative terms, which are worldsheet instantons from the point of view of the mirror dual type IIA theory. One has thus a superpotential as in \eqref{eq:Wracetrack} with $W_0=0$ and $W \to W_0$, namely
\begin{equation}
W_0 = A e^{2\pi i a \tau} + B^{2 \pi i a \tau}\,.
\end{equation}
A supersymmetric Minkowski minimum is found for $\partial_\tau W_0=0$, giving 
\begin{equation}
\tau = -\frac{1}{2\pi i (a-b)}\log\left(-\frac{a A}{b B}\right)\,,  \qquad W_0 = B\frac{a-b}{a}\left(-\frac{aA}{bB}\right)^{-\frac{b}{a-b}}\,,
\end{equation}
and recall that ${\rm Im}\,\tau=1/g_s$.

After the dilaton is stabilized, one adds to the thus obtained GVW superpotential $W_0$ non-perturbative corrections to the K\"ahler moduli. At leading order in a large volume expansion the EFT is specified by \eqref{eq:EFTKKLTKaehlersec} and one can repeat the discussion therein to show that the domain wall bound should apply, namely that there exists a BPS domain wall interpolating two different AdS vacua. Furthermore, notice that $2\pi i \tau \sim -\log W_0 \sim 2\pi {\rm Re }T$ is a somewhat reasonable approximation if $|a-b|$ is small enough.

We can now check in the explicit examples of \cite{Demirtas:2021ote,Demirtas:2021nlu} if the domain wall bound is satisfied. All that we need is the UV cutoff, which we choose to be the Kaluza-Klein scale (again in such a way that a violation of the bound with this UV cutoff would imply a violation with any other UV cutoff), and the magnitude of the AdS vacuum energy. The Kaluza-Klein scale in units of the four-dimensional Planck mass is given by $m_{\rm KK} = g_s/\mathcal{V}^{2/3}$, where $\mathcal{V}$ is the Calabi-Yau volume in string units and $g_s$ the string coupling.
This assumes isotropic internal manifold; the non-isotropic setup being discussed in the next section. Using the values of $g_s$, $\mathcal{V}$ and $L_{\rm AdS}$ given in the five explicit examples in \cite{Demirtas:2021ote,Demirtas:2021nlu}, we are thus ready to check the domain wall bound. The result is summarized in the table \ref{tab:dwbcheck1}.

\begin{table}[ht!]
\centering
\begin{tabular}{cccccc}
\toprule
$g_s$ & $\mathcal{V}$ & $m_{\rm KK}$& $L_{\text{AdS}}$ & $m_{\rm KK}^3L_{\rm AdS}$ & \text{bound} \\
\hline
$0.011$ & $945$ & $1.14 \times 10^{-4}$ & $10^{72}$ & $1.49 \times 10^{60}$&\ding{55}\\
$0.0036$ & $388.7$ & $6.76 \times 10^{-5}$ & $9.52 \times 10^{106}$ & $2.94 \times 10^{97}$&\ding{55}\\
$0.04$ & $141.4$ & $1.47 \times 10^{-3}$ & $3.16 \times 10^{28}$ & $1.01 \times 10^{20}$&\ding{55}\\
$0.05$ & $198.1$ & $1.47 \times 10^{-3}$ & $1.87 \times 10^{31}$ & $5.95 \times 10^{22}$&\ding{55}\\
$0.009$ & $4711$ & $3.2 \times 10^{-5}$ & $2.34 \times 10^{101}$ & $7.69 \times 10^{87}$&\ding{55}\\
\bottomrule
\end{tabular}
\caption{Values of $g_s$, $\mathcal{V}$ (in units of $M_s$), $m_{\text{KK}}$ (in units of $M_{\rm Pl,4}$) and $L_{\rm AdS}$ (in units of $M_{\rm Pl,4}$) for the AdS vacua in \cite{Demirtas:2021ote,Demirtas:2021nlu}. For the domain wall bound to be satisfied, the value of $m_{\rm KK}^3 L_{\rm AdS}$ should be smaller than 1 (in units of $M_{\rm Pl,4}$). The examples follow the same order of those in \cite{Demirtas:2021ote}.}
\label{tab:dwbcheck1}
\end{table}

We find that in all of the examples of \cite{Demirtas:2021ote,Demirtas:2021nlu} the domain wall bound is  violated, since $m_{\rm KK}^3 L_{\rm AdS}$ is parametrically larger than one. Intuitively, this happens because in these models the cosmological constant is exponentially small while the Kaluza-Klein scale is only polynomially so; then the domain wall bound forces us to compare the two. 
An interpretation of this violation is that the EFT of these models is inconsistent unless one integrates in the degrees of freedom of the domain wall, which are open string moduli. At this stage, it is not clear to us if the AdS vacua above will persist after introducing the new light modes. In section \ref{sec:lightdw} we will comment more extensively on this.

\subsection{KKLT models with warped throats}

KKLT-type AdS vacua can be divided into two classes: those with warped throats and those without. Warped throats occur when fields are stabilized close to a conifold point, such as in the seminal Klebanov-Strassler solution \cite{Klebanov:2000hb}. This mechanism is believed to be rather generic and its importance is in the hypothetical  uplift to de Sitter space, where the (exponential) redshift of energy scales furnished by a warped throat can potentially control and fine-tune the supersymmetry breaking uplift energy of the anti-$D3$-brane.

In presence of a warped throat there is a local tower of Kaluza-Klein modes that redshifts and this significantly lowers the EFT cutoff. Hence, domain walls that used to be lighter than $\Lambda_{\rm UV}$ can now be heavier in such a way that the KKLT-type vacua eventually satisfy the domain wall bound. In the appendix \ref{Appendix} we verify that this can indeed happen in the explicit AdS vacua constructed in \cite{McAllister:2024lnt}, while commenting on related subtleties. 

It is not clear whether this means that these vacua ultimately obey the bound with certainty. In \cite{McAllister:2024lnt}, it is suggested that the leading contributions to $W_0$ and hence the cosmological constant come from bulk fluxes. This implies that the 5-branes realizing the domain wall setting the AdS scale do not wrap 3-cycles down the throat. An equivalent choice would be to assume that the 3-form fluxes down the throat do not contribute at all. Mathematically, we can translate this into the requirement that the throat fluxes should not be of complexity type $(0,3)$. While this is certainly true in the supersymmetric Klebanov-Strassler solution, it is not clear (to us) that localized fluxes remain without $(0,3)$-piece from the bulk point of view. 
Nevertheless, let us continue by assuming that one can engineer throats with 3-form fluxes that do not contribute to $W_0$, or do not dominate. Those 3-form fluxes still determine the magnitude of the warping in the throat, thus effecting both the four-dimensional Planck mass, which is sensitive to the warped volume, and the uplift energy. Therefore, any domain wall process that changes 3-form fluxes down the throat will change the vacuum energy (which is given by $M_{\rm Pl, 4}^2L_{\rm AdS}^{-2}$) and the cosmological constant after uplift. As commented further in appendix \ref{Appendix}, this means that warped models might be equally suspicious from a domain wall viewpoint as unwarped ones, even when wrapped 5-branes down the throat do not directly affect the value of $W_0$.

\subsection{Type IIB compactifications and light domain walls}
\label{sec:lightdw}

As we have discussed, the classical Gukov-Vafa-Witten contribution $W_0$ to the superpotential in type IIB compactifications with minimal supersymmetry in four dimensions measures the tension, $T_{\rm dw}$, of the domain walls that discharge 3-form fluxes. This is the tension of $(p,q)$ 5-branes that wrap 3-cycles Poincar\'e dual to the 3-form fluxes,
\begin{equation}
T_{\rm dw}= e^{K/2}|W_0|\,.
\end{equation}
Therefore, any scenario of type IIB moduli stabilization that relies on finding small $e^{K/2}|W_0|$, such as KKLT, will be in danger of violating the domain wall bound. In this section, we propose an interpretation for what it means for the would-be EFT, if it were to exist, to violate the domain wall bound. The interpretation is really specific to the four-dimensional EFT of these type IIB compactifications, but one can possibly contemplate similar arguments in other cases.

A violation of the domain wall bound means that domain wall bubbles are light enough to be included into the EFT. These bubbles are typically microscopic, such that one can think of them as particles, but they do not need to be stable: what matters is that they are produced at energy scales well-below the cutoff. To understand how they lead to an inconsistency of the EFT, it is useful to pass from the particle picture to the field picture. 
The inconsistency will be triggered by the fact that the EFT misses a field that is light.

To pass from domain wall particles to fields, we rely on the Kachru-Pearson-Verlinde (KPV) process describing brane-flux transitions \cite{Kachru:2002gs}. This provides us with a microscopic description of the bubbles. Recall that 3-form fluxes come together with a tadpole 
\begin{equation}
0 = \int dF_5 = \int F_3 \wedge H_3 + N_{D3} \,,
\end{equation}
where $N_{D3}$ is the total charge of spacetime-filling $D3$-branes, including also the induced charge from $D7$-branes and $O3/O7$-planes.\footnote{Comparing to \cite{Giddings:2001yu}, we are denoting $N_{D3} = 2 \kappa_0^2 \,T^E_{D3}\, \,n$, with $2\kappa_0^2 = (2\pi)^7 \alpha'^4$, $T_{D3}^E=(2\pi)^{-3}(\alpha')^{-2}$ the Einstein frame $D3$-brane tension, and $n \in \mathbb{Z}$.} This tadpole implies that domain walls that change the 3-form fluxes must also change the effective number of $D3$-branes. More precisely, $N_{D3}$ receives contributions from the net number of $D3$-branes but also from the Calabi-Yau topology, providing induced $D3$-brane charge from $D7/O7/O3$-sources. We assume for simplicity that the process we are describing occurs in a given Calabi-Yau class, so that we can really think of it as changing the net number of $D3$-branes.

Following \cite{Kachru:2002gs}, let us consider an $NS5$-brane wrapping a 3-cycle which is hodge dual to a 3-cycle with $K$ units of $H_3$ flux. This furnishes the microscopic realization of our domain wall. Let us assume this wrapped 3-cycle is pierced by $M$ units of $F_3$ flux such that we have a contribution of $MK$ to the tadpole. 
To cancel the Freed-Witten anomaly \cite{Freed:1999vc}, an $NS5$-brane wrapping a 3-cycle filled with $M$ units of $F_3$ creates $M$ spacetime-filling $D3$-branes, see $e.g.$ \cite{Bouwknegt:2005ky}. Indeed, this ensures that the tadpole remains fulfilled as one moves through the domain wall.

\begin{figure}[ht!]
    \centering
    \begin{tikzpicture}[scale=0.24]

    \draw[thick, color=cyan, fill=gray!5!white] (-4,0) circle (5);
    \draw (0.25,1.0) rectangle (1.25,2.0);

    \begin{scope}
        \clip (9,-10) rectangle (29,10);
        
        \fill[color=gray!5!white] (-6,-11) circle (27);
        \draw[color=cyan] (-6,-11) circle (27);
        \draw[color=cyan] (-6,-11) circle (29);
        
        \draw[color=cyan, fill=cyan!10!white, even odd rule] (-6,-11) circle (27) (-6,-11) circle (29);
    \end{scope}
    
    \draw[ultra thin, gray] (0.25,1.0) to (9,-10);
    \draw[ultra thin, gray] (1.25,1.0) to (9,-1);
    \draw[ultra thin, gray] (1.25,2.0) to (9,3);
    \draw[ultra thin, gray] (0.25,2.0) to (9,10);
    \draw[thin, gray] (9,-10) rectangle (29,10);
    
    \draw[thin, dotted, ->] (23,2) to[out=270,in=20] (20.40,1);
    \draw (23,4) circle (2);
    \fill[darkblue] (23,6) circle (0.25);
    \fill[darkblue] (23,2) circle (0.25);

    \draw[color=magenta] (23-1.732,5) to[out=-15,in=195] (23+1.732,5) node[right,color=black]{NS5'};
    \draw[densely dotted, color=magenta] (23+1.732,5) to[out=165,in=15] (23-1.732,5);
    
    \draw[thin, dotted, ->] (15,-1) to[out=90,in=200] (18.66,0);
    \draw[fill=white] (15,-3) circle (2);
    \fill[darkblue] (15,-1) circle (0.25);
    \fill[darkblue] (15,-5) circle (0.25);

\node[font=\tiny] at (13,-0.3) {$\psi=\pi$};
\node[font=\tiny] at (13,-5.5) {$\psi=0$};
\node[font=\tiny] at (18.5,-4.5) {$\psi(\vec{x},t)$};
    
    \draw[color=magenta] (15-1.732,-4) node[left,color=black]{NS5'} to[out=-15,in=195] (15+1.732,-4);
    \draw[densely dotted, color=magenta] (15+1.732,-4) to[out=165,in=15] (15-1.732,-4);
    
    \node[align=center] at (-4,0.){$(K-1)_H$ \\ $(M+N)_{\text{D3}}$};
    \node[align=center] at (-14,2.5){$K_H$ \\ $N_{\text{D3}}$};
    \node[] at (-4,-6){NS5 wall};
    
    \end{tikzpicture}
  \caption{
On the left, the domain wall bubble (cyan) in four spacetime dimensions formed by wrapping an $NS5$-brane on an internal 3-cycle. Inside (outside of) it are $M+N$ ($N$) spacetime-filling $D3$-branes needed to cancel the Freed--Witten anomaly. On the right, zooming in on the domain wall, the polarization of the $D3$-branes into an $NS5'$-brane wrapping a contractible 2-cycle (magenta) within the 3-cycle.
}
    \label{fig:KPV}
\end{figure}
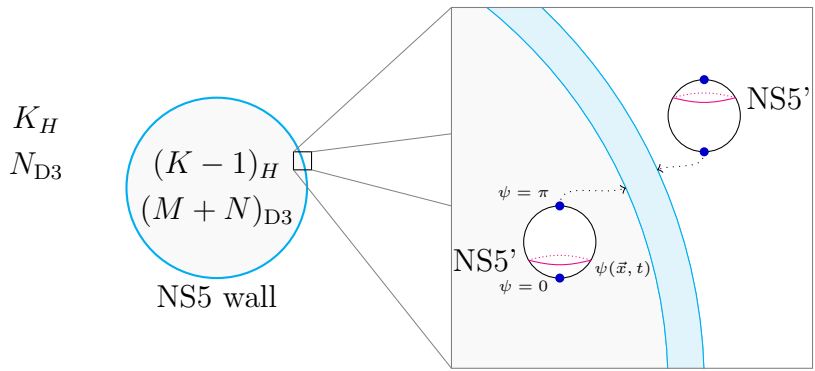

Now, to understand the origin of the sought-after light field, the key idea is that the spacetime-filling $D3$-branes polarise into an $NS5'$-brane as the domain wall is crossed (the prime notation is to distinguish it from the $NS5$-brane previously introduced).  Polarization really means that this $NS5'$-brane wraps a contractible 2-cycle inside the 3-cycle and, as one moves through the $NS5$ wall mediating the change $K\rightarrow K-1$, the $NS5'$ goes from one side of the 3-cycle to the other, where it pinches and leaves $M+N$ $D3$-branes behind. Effectively, polarization allows us to trade a non-abelian stack of $D3$-branes for a single $NS5$-brane on a contractible 2-cycle. The process is depicted in figure \ref{fig:KPV}, inspired from \cite{Shiu:2022oti}.
The ``angle" $\psi$ determining the position of the contractible $NS5'$ 2-cycle inside the 3-cycle corresponds to an axion-like scalar field in four dimensions with a monodromy.  It is the field whose particle excitations correspond to the membrane particles in four dimensions, $i.e.$ $\psi$ is the sought-after light field.

The expression for its potential energy can be found by a calculation similar to the one of \cite{Kachru:2002gs}, upon replacing anti-$D3$-branes with $D3$-branes, or equivalently upon using the appropriate $NS5'$-brane with Wess-Zumino coupling proportional to $\mathcal{F}_2 \wedge C_4$ (it is convenient to adopt a gauge in which $B_6=0$ \cite{Gautason:2016cyp}). One thus finds 
\begin{align}
\label{eq:Vkpv}
 \frac{V(\psi)}{T_{NS5} R_0^2}  \simeq   \sqrt{b_0^4\sin^4\psi \!+\! \left(\!4\pi^2g_s\!\frac{N_{D3}}{R_0^2}\!+\!\psi\!-\!\frac12\sin2\psi\!\right)^2}\!  -\!\left(\!4\pi^2g_s\frac{N_{D3}}{R_0^2} \!+\!\psi\!- \!\frac12\sin2\psi\!\right) \,,  
\end{align}
where $T_{NS5}$ is the $NS5'$-brane tension, $R_0$, $b_0$ are parameters of the configuration, and $N_{D3}$ equals the net number of mobile $D3$-branes. In particular, $b_0$ is a constant of order one, related to the specific geometry, and $R_0$ is the radius of the $A$-cycle, which is the cycle dual to the one filled with $H_3$ flux (which we are discharging).  
The square root is the DBI term and the other term is the WZ term containing ($B_6$ and) $\mathcal{F}_2\wedge C_4$. 

\begin{figure}
    \centering
    \includegraphics[width=0.65\linewidth]{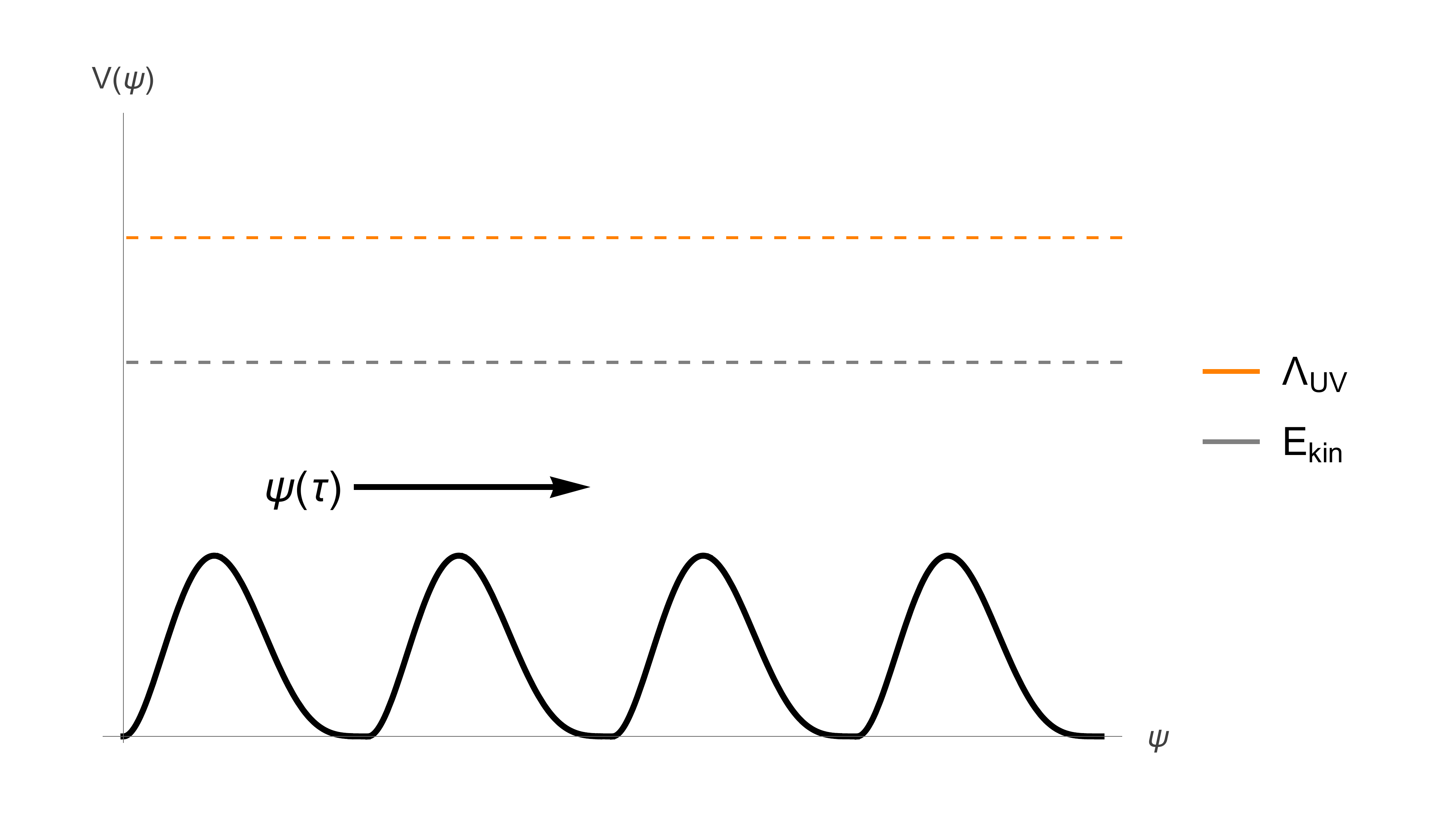}
    \caption{\small{The KPV scalar $\psi$ and its potential whose barriers are well below the cutoff scale $\Lambda_{\rm UV}$ of the EFT. Any dynamical process that dumps $E_{\rm kin}$ kinetic energy into $\psi$ that is below the cutoff but above the barrier height, makes $\psi$ ``fly over" the landscape of minima. }}
    \label{fig:potential}
\end{figure}

When the $NS5'$-brane pinches at $\psi=0$ the action collapses to that of a $D3$-brane since $4\pi^2g_sT_{{NS5}}=T_{D3}$. Recall that in an imaginary-self-dual flux background the constant terms in the DBI and WZ piece of a mobile $D3$-brane cancel as in the above expression; for anti-$D3$-branes, instead, they add up. Let us also note that the scalar potential in \cite{Kachru:2002gs} is explicitly derived  for the case in which the involved 3-cycles are the $A$- and $B$-cycles in a local Klebanov-Strassler throat, but similar expressions exist for other geometries and other branes \cite{Gautason:2015tla}. 
The potential and its various minima at $\psi=k\pi$ with $k\in \mathbb{Z}$ are shown in Figure \ref{fig:potential}.

The potential is characterized only by one scale, set by the prefactor $T_{NS5} R_0^2$, which determines all other scales. 
On dimensional grounds $\psi$ has a mass scale set by the $NS5$ domain wall tension divided by $R_0$; on top of that, one should also check the normalization of the kinetic term of $\psi$, which is not canonical; see \cite{Shiu:2022oti} for more details. However, by expanding the potential \eqref{eq:Vkpv} around $\psi=0$ one finds that $V \sim \mathcal{O}(\psi^4)$ and thus $\psi$ is actually massless on the vacuum. 
Integrating it in has far reaching consequences.

This $NS5$ domain wall interpolates between supersymmetric Minkowski minima of the type constructed  in \cite{Giddings:2001yu}.\footnote{Our arguments rest on certain simplifying assumptions, such that the 5-branes to be considered will be probes in an asymptotically flat background. We believe this can be justified in the explicit AdS vacua of \cite{Demirtas:2021nlu,McAllister:2024lnt}, whose cosmological constant is parametrically small in magnitude.} The on-shell action of a flow between two such vacua is given by \cite{Kachru:2002gs}
\begin{equation}
S = \int dtdxdy \sqrt{g_3} \,T_{\rm dw}    \qquad \text{with}\qquad T_{\rm dw} = T_{NS5}{\rm  vol}_{A}\approx T_{NS5} R_0^3\,.
\end{equation}
This is a flow associated to the $NS5'$-brane moving from $\psi=0$ to $\psi=\pi$ and we see that its on-shell action matches exactly the tension of the $NS5$-brane wrapping the whole 3-cycle $A$. Recalling that the on-shell action describes a thick domain wall, namely one which is resolved by the EFT, the above result identifies it with a thin domain wall, namely one which is fundamental (the wrapped $NS5$-brane).
Hence, by integrating in $\psi$, what was considered to be a thin (fundamental) domain wall becomes really a thick domain wall with the same tension.

Since the field $\psi$ features an axion monodromy potential with barrier heights set by the domain wall tension, any kinetic energy well below the UV cutoff that goes into $\psi$ will make the theory flow to other vacua, as depicted in Figure \ref{fig:potential}.
This means that we do not have a sensible EFT: small changes bring us away from any vacuum and it is unlikely that the classical vacuum state is meaningful, neither an EFT of fluctuations around it.

\section{Conclusions}

In this work, we provided evidence for an implementation of the covariant entropy bound \cite{Bousso:1999xy} in $d$-dimensional AdS space, $L_{\rm AdS}\Lambda_{\rm UV}^{d-1}<M^{d-2}_{\text{Pl,d}}$. 
Crucially, our argument for such an inequality relied on domain walls instead of entropy. We demanded that domain wall transitions changing the magnitude of the AdS cosmological constant are fundamental in the sense that the domain wall tension $T_{\rm dw}>\Lambda_{\rm UV}^{d-1}$. This means the domain wall cannot be resolved within the EFT, otherwise its fluctuations should be treated as light degrees of freedom
\footnote{We refer $e.g.$ to \cite{Lanza:2020qmt} for a discussion on domain wall tensions and EFT cutoff scales.}. In string theory this implies that the open string degrees of freedom living on such domain walls are so light they need to be integrated in and invalidate the EFT that has not included them. Once this is done, the same domain walls should become thick domain walls instead of thin (fundamental) domain walls.

Implementation of this domain wall bound leads to consistency constraints on AdS compactifications of higher-dimensional theories. For two-dimensional AdS vacua, it has been shown that the bound implies absence of scale separation \cite{Cribiori:2024jwq}. In this work, instead, we looked at the consequences for three- and four-dimensional AdS compactifications with scale separation. We showed that scale-separated vacua obtained from classical moduli stabilization, as well as the LVS scenario, satisfy the bound. In contrast, KKLT-like constructions may violate it, especially when their AdS vacuum energy is exponentially small. This means that domain wall transitions in these vacua are not well suppressed and control over the EFT is not obvious. More generally, the bound acts as a lower bound on the gravitino mass $e^{K/2}|W_0|$ in AdS, which so far has been not much employed in string model constructions; it also realizes the gravitino conjecture \cite{Cribiori:2021gbf,Castellano:2021yye} and the anti-de Sitter distance conjecture \cite{Lust:2019zwm} of the swampland program. Given the smallness of the observed dark energy density, lower bounds on AdS cosmological constants are an additional hurdle that needs to be tackled for realistic uplifts to de Sitter.

Suppose one considers the derivation of the KKLT minima as consistent. Then, we would have a theory with various potential minima separated by ``small" barriers and thick domain walls. In other words, the wave function of the ground state is not localized in a single classical vacuum but spreads out, and we should have a superposition of vacua with different flux numbers.\footnote{This is somewhat analogues to QCD where the actual vacuum state is the $\theta$-vacuum that is a superposition of wave functions corresponding to the classical vacua of specified winding number. } An EFT is usually derived on the grounds of having a classical vacuum and fluctuations around it. Here this picture seems altered, and it is not clear to us to what extent this affects the consistency of the construction of KKLT-like vacua we have discussed.

In recent years, several other arguments to check the consistency of AdS vacua have relied on domain walls, entropy counting and holography, see $e.g.$ \cite{Lust:2022lfc, Bena:2024are,Bobev:2023dwx,Apers:2025pon,Bobev:2025yxp}. It would be interesting to understand to what extend these ideas are related to the domain wall bound. It would be also important to improve on the approximations made in section \ref{sec:lightdw} and thus revisit the argument therein by taking into account corrections to the probe approximation and by studying the 5-branes in the appropriate instanton background.

Finally, we want to stress the non-triviality of the way the domain wall bound is encoded in the supergravity equations of motion.
For example, the CKN bound \cite{Cohen:1998zx}, which is similar to \eqref{eq:CEBd} but more restrictive, reads $\Lambda_{\rm UV}^d \leq M_{{\rm Pl}, d}^{d-2}\Lambda_{\rm IR}^2$. In our case, it may be rephrased as
\begin{equation}
\Lambda_{\rm UV}^d \leq \frac{M_{{\rm Pl,d}}^{d-2}}{L_{\rm AdS}^2}\, .
\end{equation}
Intuitively, this bound stems from the requirement that a weakly interacting system should not be a black hole. It is a natural question if also the CKN bound can be derived from the supergravity equations of motion. Clearly, we do not expect domain walls again, but perhaps some other ingredient can play an analogous role. 
However, we find it hard to identify such an ingredient. Providing a microscopic realization of the CKN bound remains thus an open problem. 

\paragraph{Acknowledgments:} We would like to thank Fien Apers, Ralph Blumenhagen, Fotis Farakos, Daniel Junghans, Andriana Makridou and Jakob Moritz for useful discussions. The work of NC is supported by the Research Foundation Flanders (FWO grant 1259125N). The work of AP was supported by a short term scientific mission grant from the COST action CA22113 THEORY-CHALLENGES. The work of TVR is supported by the KU Leuven C1 grant ZKE7799C16/25/010.

\appendix

\section{KKLT-type vacua with warped throats}\label{Appendix}

The KKLT-type geometries considered in \cite{Demirtas:2021nlu} are isotropic, while a warped throat introduces a certain degree of anisotropy. To date, the most explicit examples of KKLT-like AdS vacua including warped throats have been presented in \cite{McAllister:2024lnt}.
In this appendix, we will briefly highlight the main novelties associated to the introduction of a warped throat in the context of our domain wall bound.

In the spirit of the main text, we will assume that the EFT is consistent, this time in the warped throat, and thus proceed until we reach a contradiction. 
Determining the UV cutoff for the warped throat is non-trivial.\footnote{By making use of the emergence proposal \cite{Palti:2019pca}, in  \cite{Blumenhagen:2019qcg,Blumenhagen:2020dea} it was suggested that one should consider the mass of an anti-$D3$ brane as the universal cutoff for the EFT in the throat. Given that the emergence proposal is still actively investigated, see $e.g.$  \cite{Blumenhagen:2024lmo}, we prefer not to enter into these issues here and rather restrict ourselves to a more conservative choice explained next.} We thus focus on examining if the Kaluza-Klein modes in the warped throat satisfy the domain wall bound. Since they contribute the lightest tower in the EFT, any violations  would imply that the bound is most likely violated in the full theory.

As given in \cite{McAllister:2024lnt},  the mass scale of the Kaluza-Klein modes in the throat is given by
\begin{equation}\label{mass kkw 1}
    m_{\rm KK,w}^2\simeq\Sigma^{-1/2}\left(\frac{3}{8\pi}\right)^{1/3}\frac{|z_{\rm cf}|^{2/3}}{(g_sM^2)^{3/2}}\left(\frac{1}{\mathcal{V}_E^2\tilde{\mathcal{V}}}\right)^{1/3}\,,
\end{equation}
where $z_{\rm cf}$ is the conifold modulus, $\mathcal{V}_E=\mathcal{V}/g_s^{3/2}$ the Einstein frame volume of the Calabi-Yau, $\tilde{\mathcal{V}}$ the volume of the mirror dual in type IIA string frame, $g_s$ the string coupling, $M$ a flux quantum and the numerical factor $\Sigma\simeq1.13983$ is a parameter of the Klebanov-Strassler solution for the throat.  The AdS cosmological constant is given by
\begin{equation}
    |V_{\rm AdS}|=\frac{3}{16}\,g_s|W_0|^2\frac{1}{\mathcal{V}_E^2||\Omega||^2}\,,
\end{equation}
and, using that at large complex structure $||\Omega||^2\simeq 8\tilde{\mathcal{V}}$, we can recast \eqref{mass kkw 1} into 
\begin{equation}
    m_{\rm KK,w}^2\simeq\left( \Sigma^{-1/2}\left(\frac{16}{\pi}\right)^{1/3}\frac{|z_{\rm cf}|^{2/3}}{(g_sM^2)^{3/2}}\left(g_s|W_0|^2\right)^{-1/3}\right)|V_{\rm AdS}|^{1/3}\,.
\end{equation}
The domain wall bound  $m_{\rm KK,w}\leq |V_{\rm AdS}|^{1/6}$ reduces to the following inequality from which $|V_{\rm AdS}|$ drops out,
\begin{equation}
\label{dw bound warped KK}
\Sigma^{-1/4}\left(\frac{16}{\pi}\right)^{1/6}\frac{1}{g_s^{1/6}(g_sM^2)^{3/4}}\frac{|z_{\rm cf}|^{1/3}}{|W_0|^{1/3}}\leq1\,.
\end{equation}
This is the main relation that we will check in the explicit examples of \cite{McAllister:2024lnt}.

We start from the models termed candidate de Sitter vacua. These have AdS precursors which may morally be understood as approximate KKLT-like AdS vacua. The values of the parameters characterizing them as taken  from \cite{McAllister:2024lnt} are given in the table \ref{tab:precursor AdS vacua} below, where no violation is observed. 

\begin{table}[ht]
    \centering
\begin{tabular}{c c c c c c c}
\toprule
Id. & $z_{\mathrm{cf}}$ & $g_s$ & $M$ & $W_0$ & $m_{\rm KK,w}^3|V_{\rm AdS}|^{-1/2}$ & $\text{bound}$ \\
\midrule
1. & $1.390\times 10^{-7}$ & $0.0732$ & $16$ & $0.0103$ & $1.40\times 10^{-7}$ &
\checkmark\\
2. & $1.591\times 10^{-8}$ & $0.0595$ & $16$ & $0.0046$ & $6.26\times 10^{-8}$ &
\checkmark\\
3. & $1.121\times 10^{-7}$ & $0.0450$ & $18$ & $0.0217$ & $1.20\times 10^{-7}$&
\checkmark \\
4. & $2.369\times 10^{-6}$ & $0.0410$ & $20$ & $0.0525$ & $8.42\times 10^{-7}$&
\checkmark \\
5. & $1.468\times 10^{-6}$ & $0.0486$ & $16$ & $0.0291$ & $1.61\times 10^{-6}$&
\checkmark \\
\bottomrule
\end{tabular}
\caption{Numerical values for the conifold modulus $z_{\rm cf}$, string coupling $g_s$, flux $M$ and superpotential $W_0$ for the AdS precursors of the candidate de Sitter vacua of \cite{McAllister:2024lnt}. For the bound to be satisfied, $m_{\rm KK,w}^3|V_{\rm AdS}|^{-1/2}$ should be smaller than  1  (in units of $M_{\rm Pl,4}$). }
\label{tab:precursor AdS vacua}
\end{table}

Next, we look at other models presented in \cite{McAllister:2024lnt} which are proper AdS vacua; they are given in appendix C therein. The values of their parameters can be found in the table \ref{tab:landscapeAdS} below. Notice that the last three vacua are not supersymmetric. We see that this time some violations of the domain wall bound occur. In the following we explain the reason why some of the models above satisfy the domain wall bound and we further comment on certain technical aspects. 

\begin{table}[ht!]
\centering
\begin{tabular}{c c c c c c c}
\toprule
Id. & $z_{\text{cf}}$ & $g_{s}$ & $M$ & $W_{0}$ & $m_{\rm KK,w}^3|V_{\rm AdS}|^{-1/2}$ &$\text{bound}$ \\
\midrule
(a) & $6.0 \times 10^{-6}$  & $3.0 \times 10^{-3}$ & $8$  & $1.0 \times 10^{-35}$ & $9.18 \times 10^{32}$ &\ding{55}\\
(b) & $2.1 \times 10^{-3}$  & $1.8 \times 10^{-1}$ & $8$  & $7.4 \times 10^{-18}$ & $5.6 \times 10^{12}$ &\ding{55}\\
(c) & $2.4 \times 10^{-47}$ & $1.5 \times 10^{-2}$ & $2$  & $1.6 \times 10^{-27}$ & $1.41 \times 10^{-16}$ &\checkmark\\
(d) & $1.3 \times 10^{-42}$ & $2.7 \times 10^{-1}$ & $2$  & $3.2 \times 10^{-25}$ & $1.34 \times 10^{-17}$ &\checkmark\\
(e) & $9.1 \times 10^{-7}$  & $7.5 \times 10^{-2}$ & $14$ & $3.2 \times 10^{-2}$  & $5.02 \times 10^{-7}$ &\checkmark \\
(f) & $2.1 \times 10^{-3}$  & $1.8 \times 10^{-1}$ & $8$  & $9.7 \times 10^{-18}$ & $4.23 \times 10^{12}$&\ding{55} \\
(g) & $3.1 \times 10^{-7}$  & $2.3 \times 10^{-1}$ & $8$  & $2.3 \times 10^{-2}$  & $1.35 \times 10^{-7}$&\checkmark \\
(h) & $1.0 \times 10^{-22}$ & $5.6 \times 10^{-2}$ & $4$  & $1.9 \times 10^{-11}$ & $5.82 \times 10^{-11}$&\checkmark \\
(i) & $2.0 \times 10^{-5}$  & $1.5 \times 10^{-1}$ & $14$ & $2.9 \times 10^{-1}$  & $1.81 \times 10^{-7}$&\checkmark \\
(j) & $1.1 \times 10^{-10}$ & $5.9 \times 10^{-2}$ & $14$ & $7.9 \times 10^{-4}$  & $4.75 \times 10^{-9}$ &\checkmark\\
(k) & $2.5 \times 10^{-7}$  & $7.9 \times 10^{-2}$ & $14$ & $5.9 \times 10^{-2}$  & $6.48 \times 10^{-8}$&\checkmark \\
\bottomrule
\end{tabular}
\caption{Numerical values for the conifold modulus $z_{\rm cf}$, string coupling $g_s$, flux $M$ and superpotential $W_0$ 
for the AdS vacua in appendix C of \cite{McAllister:2024lnt}. For the domain wall to be satisfied, the value of $m_{\rm KK,w}^3|V_{\rm AdS}|^{-1/2}$ should be smaller than  1  (in units of $M_{\rm Pl,4}$).}
\label{tab:landscapeAdS}
\end{table}

The (precursors to the) candidate de Sitter vacua are selected with, among other requirements, a certain property called alignment. This condition effectively asks for the AdS cosmological constant to be roughly of the same order as the potential of an anti-D3-brane, 
\begin{equation}
\Xi =\frac{V_{\overline{D3}}}{|V_{\rm AdS}|}\simeq \frac{|z_{cf}|^{4/3}}{|W_0|^2}\frac{\mathcal{V}_E^{2/3}\tilde{\mathcal{V}}^{1/3}}{(g_s M)^2}\zeta\simeq 1 \, ,
\end{equation}
with $\zeta\simeq 114$  and where $V_{\overline{D3}} = c/\mathcal{V}_E^{\frac43}$, with $c=\eta \frac{z_{cf}^{4/3}}{g_s M^2 \tilde{\mathcal{V}}^{2/3}}$, $\eta \simeq 2.6727$ \cite{McAllister:2024lnt}.
It is argued in \cite{McAllister:2024lnt} that this ensures an uplift mechanism avoiding runaway solutions. We can recast this condition into the form
\begin{equation}
\label{eq:alignmentcond}
\frac{1}{g_s^{1/6}(g_s M^2)^{3/4} }\frac{|z_{\rm cf}|^{1/3}}{|W_0|^{1/3}}\simeq \Xi^{1/4}|V_{\rm AdS}|^{1/12} (g_s M^2)^{-1/2}\zeta^{-1/4}\,.
\end{equation}
Then, by comparing it with \eqref{dw bound warped KK}, we see that for the domain wall bound to be satisfied, the right hand side of \eqref{eq:alignmentcond} must not be (much) larger than one. Since $\zeta^{-1/4}\simeq 0.3$, while typically in these solutions $g_s M^2 \gg 1$ and $|V_{\rm AdS}| \ll 1$, we see that the key parameter is really $\Xi$. If $\Xi \gg1 $, the domain wall bound is likely violated, while for $\Xi \lesssim 1$, as forced by the alignment condition, the bound should be safe. This is indeed what we observe in the explicit examples above, where the three AdS vacua violating the bound in table \ref{tab:landscapeAdS} have alignment parameters respectively of order $10^{70}$, $10^{34}$, and $10^{35}$.

The role of the alignment condition in the AdS EFT can be elaborated further.
Let us assume the throat description is valid up to some scale $\Lambda$. For the uplift to be describable within the throat EFT, one would demand that $\Lambda>V_{\overline{D3}}^{1/4}=\Xi^{1/4}|V_{\rm AdS}|^{1/4}$. At the same time, the domain wall giving rise to the AdS cosmological constant should certainly not be within the same description, \textit{i.e.} $\Lambda<T_{\rm dw}^{1/3}\simeq|V_{\rm AdS}|^{1/6}$.
The two conditions combine into $\Xi^{1/4}|V_{\rm AdS}|^{1/4}<|V_{\rm AdS}|^{1/6}$, which is safe for  $\Xi\sim 1$, but can fail for $\Xi\gg 1$.

The requirement $\Lambda>\Xi^{1/4}|V_{\rm AdS}|^{1/4}$ can also be more concretely employed to estimate the quality of the approximation of the UV cutoff with the warped Kaluza-Klein scale. Indeed, from the ratio
\begin{equation}
\label{comparison KK uplift}
\frac{m_{\rm KK,w}^2}{|V_{\rm AdS}|^{1/2}}\simeq\left[ \Sigma^{-1/2}\left(\frac{16}{\pi}\right)^{1/3}\frac{|z_{\rm cf}|^{2/3}}{(g_sM^2)^{3/2}}\left(g_s|W_0|^2\right)^{-1/3}\right]|V_{\rm AdS}|^{-1/6}\,,
\end{equation}
we can determine if $m_{\rm KK,w}$ is below the uplift scale $V_{\overline{D3}}^{1/4} = \Xi^{1/4}| V_{\rm AdS}|^{1/4}$ for vacua satisfying the alignment condition $\Xi \sim 1$. Looking at the five precursors to the candidate de Sitter vacua of \cite{McAllister:2024lnt}, and using the de Sitter cosmological constant as proxy for $|V_{\rm AdS}|$ (the error being a multiplicative order one factor) we find the values displayed in table \ref{mkk uplift comparison} below. 
\begin{table}[ht]
    \centering
\begin{tabular}{c c c c c c}
\toprule
Id. & $z_{\mathrm{cf}}$ & $g_s$ & $M$ & $W_0$ & $m_{\rm KK,w}|V_{\rm AdS}|^{-1/4}$\\
\midrule
1. & $1.390\times 10^{-7}$ & $0.0732$ & $16$ & $0.0103$  & 0.19 \\
2. & $7.934\times 10^{-9}$ & $0.0595$ & $16$ & $0.0046$  & 0.18 \\
3. & $1.121\times 10^{-7}$ & $0.0450$ & $18$ & $0.0217$ &  0.17\\
4. & $2.369\times 10^{-6}$ & $0.0410$ & $20$ & $0.0525$ & 0.16\\
5. & $1.468\times 10^{-6}$ & $0.0486$ & $16$ & $0.0304$ & 0.20\\
\bottomrule
\end{tabular}
\caption{Comparison of the Kaluza-Klein to the uplift scale for the candidate vacua of \cite{McAllister:2024lnt}.}\label{mkk uplift comparison}
\end{table}
We see that the warped Kaluza-Klein scale in these models is below the uplift scale, thus confirming that it is underestimating the UV cutoff. 
Observe however that the domain wall bound estimate is quite sensitive to the UV cutoff; $e.g.$ considering $10^2 m_{\rm KK,w}$ instead of $m_{\rm KK,w}$ would lead to a violation of the bound for the last of the candidate vacua. These estimates highlight how the restrictions coming from the domain wall bound are competing with $\Lambda > V_{\overline{D3}}^{1/4}$.

One could consider examples with even stronger redshifting, hoping to establish the desired hierarchies while complying with the bound. Importantly, this question is related to the likelihood of constructing vacua with a large amount of flux in the throat, $Q_{\rm throat}^{\rm flux}$, in the presence of $n_{\rm cf}$ conifold regions. Those are the most promising for providing setups compatible with the domain wall bound, due to the lowering effect of $Q_{\rm throat}^{\rm flux}$ on the value of the conifold modulus \cite{McAllister:2024lnt}
\begin{equation}
    |z_{\rm cf}|=\frac{1}{2\pi}e^{-\frac{2\pi}{g_sM^2\, n_{\rm cf}}Q_{\rm throat}^{\rm flux}}\,.
\end{equation}
It is intriguing that throats containing almost the entire $D3$-brane charge were found to be exponentially rare in \cite{McAllister:2024lnt}, but a more detailed analysis falls beyond the scope of this paper.
We hope to return to these issues in the future.

Let us also remind that the solutions we have considered have been constructed with $g_sM\simeq 1$, which is on the borderline of parametric control 
\cite{Blumenhagen:2019qcg}. Naively looking at \eqref{dw bound warped KK}, one might consider that increasing the values of $g_sM^2$ would allow for much smaller values of $e^{K/2}|W_0|$. However, in \cite{Blumenhagen:2019qcg,Blumenhagen:2020dea,Blumenhagen:2022dbo} the cutoff of the EFT in the throat was found to be given by $\Lambda\simeq (g_sM^2)\,m_{\rm KK,w}$. Inserting this into \eqref{dw bound warped KK} we get
\begin{equation}
    \frac{(g_sM^2)^{1/4}}{g_s^{1/6}}\frac{|z_{\rm cf}|^{1/3}}{|W_0|^{1/3}}\leq 1\,,
\end{equation}
which seems generically hard to satisfy for  small $g_s$ and large $g_sM^2$.

To summarize, our findings indicate that there may be room for suspicion also for the AdS vacua with warped throats. Assuming that $W_0$ is dominated by bulk fluxes, as suggested in \cite{McAllister:2024lnt}, we would expect that our bound should be satisfied much more comfortably than what we actually find, as it compares the lightest redshifted tower with very heavy domain walls in the bulk. If present, redshifted domain walls would lie close to warped Kaluza-Klein scale, bringing the model back to the shortcomings of the unwarped setup. Additionally, if we also consider a potential uplift to de Sitter, regardless of whether we have bulk or throat domain walls, to avoid violations of the bound their tension should be comparable to that of the anti-$D3$-brane, as the previous discussion on the alignment condition suggests. This leads to a paradox: on the one hand the anti-$D3$-brane uplift energy should not modify the setup too much, on the other hand at the domain wall energy scale the system is considerably modified; but we have just argued that these two scales are comparable.

\bibliography{references}  
\bibliographystyle{utphys}

\end{document}